\documentclass[%
12pt,
reprint,
superscriptaddress,
preprintnumbers,
amsmath,amssymb,
aps,
prb,
floatfix,
]{revtex4-2}

\usepackage{graphicx}
\usepackage{rotating}
\usepackage{dcolumn}
\usepackage{bm}
\usepackage[colorlinks, allcolors=blue]{hyperref}

\usepackage{orcidlink}

\usepackage{xcolor}
\usepackage{microtype}
\usepackage{graphicx}
\usepackage[multidot]{grffile}

\usepackage{tikz}
\usetikzlibrary{shapes, arrows}
\tikzstyle{line} = [draw, -latex']
\usepackage[defaultcolor=red]{changes}

\usepackage[subpreambles=true]{standalone}

\begin{document}

\title{
\textit{Ab initio} calculations of third-order elastic coefficients
}
\author{Chenxing Luo\,\orcidlink{0000-0003-4116-6851}}
\affiliation{Department of Applied Physics and Applied Mathematics, Columbia University, New York, NY 10027, USA}

\author{Jeroen Tromp\,\orcidlink{0000-0002-2742-8299}}
\email[]{jtromp@princeton.edu}
\affiliation{Princeton University, Princeton, NJ 08544, USA}

\author{Renata M. Wentzcovitch\,\orcidlink{0000-0001-5663-9426}}
\email[]{rmw2150@columbia.edu}
\affiliation{Department of Applied Physics and Applied Mathematics, Columbia University, New York, NY 10027, USA}
\affiliation{Department of Earth and Environmental Sciences, Columbia University, New York, NY 10027, USA}
\affiliation{Lamont–Doherty Earth Observatory, Columbia University, Palisades, NY 10964, USA}

\date{\today}

\begin{abstract}

Third-order elasticity (TOE) theory predicts strain-induced changes in second-order elastic coefficients (SOECs) and can model elastic wave propagation in stressed media. Although third-order elastic tensors have been determined based on first principles in previous studies, their current definition is based on an expansion of thermodynamic energy in terms of the Lagrangian strain near the natural, or zero pressure, reference state. This definition is inconvenient for predictions of SOECs under significant initial stresses. Therefore, when TOE theory is necessary to study the strain dependence of elasticity, the seismological community has resorted to an empirical version of the theory.

This study reviews the thermodynamic definition of the third-order elastic tensor and proposes using an ``effective'' third-order elastic tensor. An explicit expression for the effective third-order elastic tensor is given and verified. We extend the \textit{ab initio} approach to calculate third-order elastic tensors under finite pressure and apply it to two cubic systems, namely, NaCl and MgO. As applications and validations, we evaluate (a) strain-induced changes in SOECs and (b) pressure derivatives of SOECs based on \textit{ab initio} calculations. Good agreement between third-order elasticity-based predictions and numerically calculated values confirms the validity of our theory.

\end{abstract}

\maketitle

\section{Introduction}

In seismological applications, second-order elastic coefficients (SOECs) are treated as a function of the initial configuration of a solid in equilibrium under initial stress \citep[e.g.,][]{dahlenElasticVelocityAnisotropy1972,dahlenTheoreticalGlobalSeismology1998}. Although studies of SOECs vs.\ pressure take into account first-order effects of hydrostatic stress \citep[e.g.,][]{wentzcovitchFirstPrinciplesQuasiharmonic2010}, addressing the effects of non-hydrostatic or \emph{deviatoric\/} stress on elasticity (i.e., stress-dependent elasticity) is equally essential owing to the non-hydrostatic nature of stress in many geological processes (e.g., plate tectonics and mantle convection) \cite{trompEffectsInducedStress2018}.
A better understanding of the effects of stress on elastic parameters also benefits many other applications involving monitoring seismic wave speeds (e.g., hydrocarbon reservoir characterization, stress formation monitoring, and volcano monitoring).

Third-order elasticity (TOE) theory is a viable approach for addressing the stress dependence of elasticity \cite[e.g.,][]{sripanichStressdependentElasticityWave2021}.
Since SOECs are generally reported under hydrostatic pressure, this condition is assumed for the initial configuration. The non-hydrostatic or deviatoric part of the stress induces a minor elastic strain away from the initial configuration \citep{dahlenElasticVelocityAnisotropy1972}.
Since third-order elastic coefficients (TOECs) describe the effects of such strains on SOECs according to a theoretically derived linear-approximated expression \citep{thurstonThirdOrderElastic1966, truesdellMechanicsSolidsVolume1984}, TOECs enable the full SOEC tensor under non-hydrostatic stress to be determined. Several variants of TOE theory have been adopted by seismologists  \citep[e.g.,][]{sinhaStressInducedAzimuthal1996, prioulNonlinearRockPhysics2004, fuckAnalysisSymmetryStressed2009}.

However, although TOE theory was explored as early as the 1960s in the ultrasound community \cite[e.g.,][]{thurstonCalculationLatticeParameter1967, bruggerThermodynamicDefinitionHigher1964}, an empirical version of TOE theory has been adopted by seismologists \citep[e.g.,][]{sinhaStressInducedAzimuthal1996, prioulNonlinearRockPhysics2004, fuckAnalysisSymmetryStressed2009}.
The reason is that TOECs are generally determined in terms of their original thermodynamic definition as third-order Lagrangian strain derivatives of the thermodynamic energy density, thereby defining the \emph{thermodynamic\/} TOECs \cite{bruggerThermodynamicDefinitionHigher1964}.
As shown in this study, the use of thermodynamic TOECs complicates the evaluation of strain effects on SOECs by requiring SOEC tensors to be carefully pulled back to a common reference frame.
Without this practice, tensors parameterized based on nonlinear rock physics modeling \citep[e.g.,][]{prioulNonlinearRockPhysics2004} are not equivalent to those based on thermodynamic TOECs.
Early developments of first-principle TOE theory invoked the natural (i.e., 0~GPa) reference frame \citep[e.g.,][]{thurstonCalculationLatticeParameter1967} as the common frame, thereby hindering its application in the multi-MPa regime \cite{dahlenElasticVelocityAnisotropy1972}.
In the ultrasound community, second- and third-order elasticity theories were developed in tandem based on Lagrangian strain derivatives of the thermodynamic energy density.
A source of challenge and confusion in geophysics is that there are at least two kinds of SOECs. \emph{Thermodynamic\/} SOECs are defined as the second-derivative of the thermodynamic energy with respect to the Lagrangian strain \cite{bruggerThermodynamicDefinitionHigher1964, thurstonEffectiveElasticCoefficients1965, truesdellMechanicsSolidsVolume1984, wallaceThermoelasticityStressedMaterials1967, dahlenTheoreticalGlobalSeismology1998, levitasNonlinearElasticityPrestressed2021},
whereas \emph{effective\/} SOECs are defined based on an incremental version of Hooke's law under initially hydrostatic conditions \cite{barronSecondorderElasticConstants1965, wallaceThermoelasticityStressedMaterials1967, thurstonEffectiveElasticCoefficients1965, truesdellMechanicsSolidsVolume1984, dahlenTheoreticalGlobalSeismology1998, levitasNonlinearElasticityPrestressed2021}; their values differ except under zero initial pressure conditions.
The lack of a complementary definition of effective TOECs explains why no first-principle TOE theory for high-pressure applications has been adopted.
The different variants of the elastic tensors mentioned above have been recently reviewed in great detail in \cite{levitasNonlinearElasticityPrestressed2021}.

In materials science, natural state (0~GPa) elastic constants up to higher (fourth or fifth) order also contain complete information about solids.
Since they can be used to determine SOECs and TOECs under finite pressure \cite{levitasNonlinearElasticityPrestressed2021}, they offer alternative pathways to address SOECs under stress.
Fourth- or fifth-order elastic constants have been determined from both the DFT \cite[e.g.,][]{chenFifthdegreeElasticEnergy2020} and planar compression experiments \cite[e.g.,][]{claytonSHOCKCOMPRESSIONMETAL2014, claytonFiniteStrainAnalysis2014}, but generally for high-symmetry systems only.
They are less practical and indirect in addressing challenges in seismic measurements than the formalism presented in this study.

This study demonstrates how to evaluate strain effects on SOEC based on TOE theory using \textit{ab initio} calculations.
The development of \textit{ab initio}-based methods to compute TOEC is an active research area \citep[e.g.,][]{zhaoFirstprinciplesCalculationsSecond2007, caoFirstPrinciplesCalculationThirdOrder2018, liaoElastic3rdToolCalculating2021, wangInitioCalculationsSecond2009, guHighPressureThirdOrderElastic2019, liaoHighefficientStrainstressMethod2022}.
We adopt the favored approach, which expands the strain energy vs.\ the Lagrangian strain \cite[e.g.,][]{zhaoFirstprinciplesCalculationsSecond2007, liaoElastic3rdToolCalculating2021, wangInitioCalculationsSecond2009, guHighPressureThirdOrderElastic2019}. Because these methods are generally developed for 0~GPa elastic coefficients, we extend and test them for SOECs and TOECs under finite pressure.

A particular case of the effects of stress on SOECs involves their pressure derivatives,
and such derivatives can be analytically expressed in terms of TOECs \citep[see, e.g.,][]{birchFiniteElasticStrain1947, barschAdiabaticIsothermalIntermediate1967, changNonlinearPressureDependence1967, truesdellMechanicsSolidsVolume1984}.
This could reasonably explain why stress-induced changes in SOECs can be conveniently described in terms of pressure derivatives of SOECs, as shown in recent studies \cite{trompEffectsInducedStress2018, trompEffectsInducedStress2019, maitraStressDependenceElastic2021}.
In this study, we validate relationships between pressure derivatives of SOECs and predictions based on TOECs; a simplified expression thanks to our introduction of effective TOECs is also validated. These effective TOECs also benefit predictions of finite-pressure elasticity based on SOEC pressure derivatives \citep[e.g.,][]{liaoPressureTemperatureDependence2021}.
The use of TOE theory to evaluate pressure derivatives of SOECs and its application to assessing strain effects on SOECs also serves as a self-consistent validation of the \textit{ab initio} approach for computing TOECs under finite pressure.

A recent study by \citet{maitraStressDependenceElastic2021} took a different approach to derive expressions for effective TOECs and the pressure derivatives of SOECs. As we will show later in the discussion, their results are similar to ours.

Given the close relationship between TOE and crystal anharmonic theory \citep[][Section~29.1]{truesdellMechanicsSolidsVolume1984}, 
a better understanding of TOE theory allows us to better address thermoelasticity or thermal expansivity \cite{carrierQuasiharmonicElasticConstants2008,liaoPressureTemperatureDependence2021} at finite pressure for highly-anharmonic (e.g., \cite{luoInitioInvestigationHbond2022}) or highly-anisotropic solids (e.g., serpentine \cite{dengElasticityLizarditeHigh2021}).

The structure of this paper is as follows. Section~\ref{sec:2} reviews the relevant theory for elastic coefficients under finite pressure and introduces the effective TOECs. Section~\ref{sec:3} computes the elastic coefficients under finite pressure. As applications and validations, Sections~\ref{sec:4} and~\ref{sec:5} evaluate strain effects on SOECs and pressure derivatives of SOECs based on our proposed theories and calculated elastic coefficients. Section~\ref{sec:conclusion} presents our conclusions.

For reference and clarity,
we summarize the notations used in different studies in Table~\ref{tab:1}.


\section{Formulation of elastic coefficients under finite pressure}
\label{sec:2}

\subsection{Reference frames and deformation}

\begin{figure}[htp]
    \centering
    


\begin{tikzpicture}
\tikzstyle{every node}=[font=\small]
\node (n0) [draw, label = {$P = 0$ GPa} ] {Natural ($a_i$)};
\node (s0) [draw, label = {$T^0_{ij}$, $A^0_{ijkl}$}, right of=n0, right=3.5em] {Initial ($X_i$)};
\node (s1) [draw, label = {$T^\mathrm{L}_{ij}$, $A^\mathrm{L}_{ijkl}$}, right of=s0, right=3.5em] {Present ($x_i$)};
\path [line] (n0) --  (s0);
\path [line] (s0) -- node [anchor=south] {$F_{ij}$} node [anchor=north] {$\eta_{ij}$} (s1);
\end{tikzpicture}

    \caption{Natural, initial, and present reference frames.
    In the natural frame at 0~GPa, we use a set of natural coordinates labeled~$\{a_i\}$\,.
    Under finite stress~$T^0_{ij}$ in the initial frame, we use a set of initial coordinates labeled~$\{X_i\}$\,.
    The thermodynamic SOECs in this frame are denoted by~$A^0_{ijkl}$\,.
    After application of a deformation induced by the deformation gradient~$F_{ij}$\,, with corresponding Lagrangian strain~$\eta_{ij}$\,,
    we reach the present reference frame with a set of present coordinates~$\{x_i\}$\,, in which the Lagrangian description of the Cauchy stress is given by~$T^\mathrm{L}_{ij}$ and the thermodynamic SOECs are denoted by~$A^\mathrm{L}_{ijkl}$\,.
    }
    \label{fig:frames}
\end{figure}
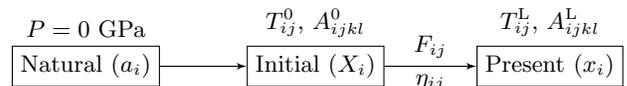

We first clarify the different kinds of reference frames commonly used to address elasticity at finite pressure. In previous studies, focusing on static properties only, there are generally three kinds of frames \cite{thurstonEffectiveElasticCoefficients1965, truesdellMechanicsSolidsVolume1984,dahlenElasticVelocityAnisotropy1972},
namely, (a) a natural frame, the 0~GPa state, (b) an initial frame, where the elastic coefficients are being evaluated, usually a hydrostatically prestressed state, and (c) the present frame, where a small deformation is applied upon the initial frame to help evaluate the curvature of the potential energy surface at the initial frame. These frames are summarized in Fig.~\ref{fig:frames}.
In previous studies (e.g., \cite{thurstonEffectiveElasticCoefficients1965, thurstonCalculationLatticeParameter1967, truesdellMechanicsSolidsVolume1984}), the natural frame serves as a common frame of reference to pull tensors back to. But to study elasticity under multi-GPa pressures, it is pointless to keep transferring elastic tensors to 0~GPa to evaluate stress effects and then transferring them back. Therefore, our subsequent discussion will focus on the initial and present frames of reference.

The deformation gradient~$F_{ri}$ relates the initial coordinates~$\{X_i\}$ and the present coordinates~$\{x_i\}$,  via
\begin{equation}  
    F_{ri} = \frac{\partial x_r}{\partial X_i} \,,
\end{equation}
and the corresponding Lagrangian strain $\eta_{ij}$ is defined by
\begin{equation}
\eta_{ij}
= \tfrac{1}{2} (F_{ki} F_{kj} - \delta_{ij})
= \tfrac{1}{2} \left(\frac{\partial x_k}{\partial X_i} \frac{\partial x_k}{\partial X_j} - \delta_{ij} \right) \,.
\end{equation}
The Jacobian~$J$ relating the volume or density in the initial state ($V_0$, $\rho_0$) to the present state ($V$, $\rho$) frame is defined as
\begin{equation}
    J = \det \mathbf{F} = V / V_0 = \rho_0 / \rho\,.
\end{equation}

\subsection{Thermodynamic definition of elastic constants}


Suppose the system has an initial volume $V_0$ and initial stress $T^0_{ij}$\,,
corresponding to the initial state in Fig.~\ref{fig:frames}. If we expand the thermodynamic energy $E$
near the initial state in powers of the Lagrangian strain $\eta_{ij}$, that is,
\begin{equation}
\begin{split}
\frac{E(T_{ij}^0, \eta_{ij})}{V_0} &=
\frac{E(T_{ij}^0, 0)}{V_0}
+ T_{ij}^0\, \eta_{ij}
+ \tfrac{1}{2!}\, A_{ijkl}^0\, \eta_{ij}\, \eta_{kl} \\
& + \tfrac{1}{3!}\, A_{ijklmn}^0\, \eta_{ij}\, \eta_{kl}\, \eta_{mn} + O(\eta^4)\,,
\end{split}
\label{eq:4}
\end{equation}
then the first-, second-, and third-order expansion coefficients are the initial stress $T_{ij}^0$\,, the thermodynamic elastic coefficients SOECs $A_{ijkl}^0$ (denoted as $\Xi_{ijkl}$ in~\cite{dahlenTheoreticalGlobalSeismology1998})\,,
and the TOECs $A_{ijklmn}^0$\,. The expansion of the Helmholtz free energy gives isothermal elastic coefficients, the expansion of the internal energy gives adiabatic elastic coefficients \cite{wallaceThermoelasticityStressedMaterials1967}, and the expansion of static (clamped ions) energies gives static coefficients (0~K but without zero-point-motion energy effects). These tensors are all evaluated in the initial state where the strain $\eta_{ij} = 0$, hence the superscript~$0$\,. The strain, stress, SOEC, and TOEC tensors used in this study are based on traditional fixed Cartesian basis vectors.
These expansion coefficients can alternatively be expressed as partial derivatives of the Helmholtz free energy with respect to the Lagrangian strain, that is,
\begin{subequations}
\begin{align}
T_{ij}^0 &= \frac{1}{V_0} \frac{\partial E}{\partial \eta_{ij}}\,,
\label{eq:5-1} \\
A_{ijkl}^0 &=\frac{1}{V_0} \frac{\partial^2 E}{\partial \eta_{ij}\, \partial \eta_{kl}}\,,
\label{eq:5-2} \\
A_{ijklmn}^0 &= \frac{1}{V_0} \frac{\partial^3 E}{\partial \eta_{ij}\, \partial \eta_{kl}\, \partial \eta_{mn}} \,,
\label{eq:5-3}
\end{align}
\end{subequations}
in accordance with the original definition of high-order elastic constants by \citet{bruggerThermodynamicDefinitionHigher1964}.

These stress and thermodynamic elastic tensors may be pulled-back or pushed-forward between different reference frames. For example, pulling the Lagrangian description of the Cauchy stress ($T^\mathrm{L}$, also known as the Cauchy stress, the Lagrangian-Cauchy stress, or the true stress) back from the present frame to the initial frame as the second Piola-Kirchhoff stress ($T^\mathrm{SK}_{ij}$) is achieved via the Piola transformation \cite{thurstonCalculationLatticeParameter1967, dahlenTheoreticalGlobalSeismology1998, maitraStressDependenceElastic2021},
\begin{equation}
T^\mathrm{SK}_{ij} = J \, F^{-1}_{ir} F^{-1}_{js}\, T^\mathrm{L}_{rs} = J \, \frac{\partial X_i}{\partial x_r} \frac{\partial X_j}{\partial x_s} \,T^\mathrm{L}_{rs} \,.
\label{eq:6}
\end{equation}
Similarly,
suppose $A^\mathrm{L}$ denotes the regular thermodynamic SOECs, whereas $A^\mathrm{SK}$ denotes its pull-back to the initial frame.
Then we have the relationship \cite{thurstonCalculationLatticeParameter1967, maitraStressDependenceElastic2021, levitasNonlinearElasticityPrestressed2021}
\begin{equation}
\begin{split}
A_{ijkl}^\mathrm{SK}  & = J \, F^{-1}_{ir} F^{-1}_{js} F^{-1}_{kp} F^{-1}_{lq}\, A_{rspq}^\mathrm{L} \\
& = J \, \frac{\partial X_i}{\partial x_r} \frac{\partial X_j}{\partial x_s} \frac{\partial X_k}{\partial x_p} \frac{\partial X_l}{\partial x_q} \,A_{rspq}^\mathrm{L}\, .
\end{split}
\end{equation}

This set of transformations brings tensors to a common frame and makes it convenient to consider higher-order strain derivatives of these tensors, for example \cite{thurstonCalculationLatticeParameter1967} (Eq.~(34)),
\begin{subequations}
\begin{align}
    A^0_{ijkl} = \partial T^\mathrm{SK}_{ij} / \partial \eta_{kl}
    \quad&\text{or}\quad
    T^\mathrm{SK1}_{ij} = A^0_{ijkl} \,\eta_{kl}\,, \label{eq:7-1} \\
    A^0_{ijklmn} = \partial A^\mathrm{SK}_{ijkl} / \partial \eta_{mn}
    \quad&\text{or}\quad
    \Delta A^\mathrm{SK}_{ijkl} = A^0_{ijklmn} \,\eta_{mn}\,, \label{eq:7-2}
\end{align}
\label{eq:7}
\end{subequations}
where the incremental second Piola-Kirchhoff stress $T_{ij}^\mathrm{SK1}$ is defined as $T_{ij}^\mathrm{SK1} = T_{ij}^\mathrm{SK} - T^\mathrm{0}_{ij}$, and where $\Delta A^\mathrm{SK}_{ijkl}$ is defined similarly, namely,
\begin{equation}
    \Delta A_{ijkl}^\mathrm{SK} \equiv  A_{ijkl}^\mathrm{SK} - A_{ijkl}^0\,. 
    \label{eq:9-1}
\end{equation}
These two equations give us the stress vs.\ strain and SOECs vs.\ strain relationships within a single frame.

\subsection{Constitutive relations and effective elastic tensors}

It might be mathematically convenient to have all tensors live in the same reference frame; however, this is no longer so in practice.


In the absence of initial stress, Hooke's law takes the form $T_{ij} = C_{ijkl} \,\epsilon_{kl}$, where $\epsilon_{kl}$ denotes the symmetric infinitesimal strain tensor which is related to the deformation tensor $F_{ij}$ by
\begin{equation}
    \epsilon_{ij} = \tfrac{1}{2} (F_{ij} + F_{ji}) - \delta_{ij} \approx \eta_{ij}\, ,
\end{equation}
which approximates the Lagrangian strain.

The presence of initial stress modifies the constitutive relationship and complicates the linearized version of the stress vs.\ strain relationship. To calculate the induced effect on stress by a strain, one needs to (a) use an ``effective'' SOEC tensor that may lack the familiar symmetries (e.g., $B_{ijkl}$ in \cite{wallaceThermoelasticityStressedMaterials1967} or $\Upsilon_{ijkl}$ in \cite{dahlenTheoreticalGlobalSeismology1998}, where $B_{ijkl}$ does not satisfy $B_{ijkl} = B_{klij}$ except under hydrostatic stress), or (b) use a symmetric ``effective'' SOEC tensor ($\Gamma_{ijkl}$ in \cite{dahlenTheoreticalGlobalSeismology1998}) that has the desired symmetries but modifies the stress vs.\ strain relationship with contributions from the deviatoric stress \cite[see][Eq.~(3.144)]{dahlenTheoreticalGlobalSeismology1998}.
Either way, an ``effective'' SOEC tensor is involved.

To facilitate a subsequent discussion on TOECs, we follow the formulation in \cite{wallaceThermoelasticityStressedMaterials1967, barronSecondorderElasticConstants1965}. 
The relationship between the symmetric incremental Lagrangian description of the Cauchy stress  $T_{ij}^\mathrm{L1}$ and the infinitesimal symmetric strain can be given in a familiar linearized form, namely,
\begin{equation}
\begin{split}
& T_{ij}^\mathrm{L1} = T_{ij}^\mathrm{L} - T_{ij}^0 = C^0_{ijkl} \,\epsilon_{kl} \\
    & \qquad \text{ or \quad}
    C_{ijkl}^0 = \partial T^\mathrm{L}_{ij} / \partial \epsilon_{kl}\, .
\end{split}
\label{eq:8}
\end{equation}
The same would be valid in the formulation in \cite{dahlenTheoreticalGlobalSeismology1998} in the absence of an initial deviatoric stress.
Here, $C^0_{ijkl}$ denote the elements of the ``effective'' elastic tensor, sometimes also known as the Wallace moduli \cite{wallaceThermoelasticityStressedMaterials1967}.
The $C^0_{ijkl}$ are related to the thermodynamic SOECs $A^0_{ijkl}$ via \cite{barronSecondorderElasticConstants1965, thurstonEffectiveElasticCoefficients1965}
\begin{equation}
\begin{split}
C^0_{ijkl} = &~ A^0_{ijkl} - T^0_{ij}\, \delta_{kl} \\
    & + \tfrac{1}{2}(T^0_{ik}\, \delta_{jl} + T^0_{kj}\, \delta_{il} + T^0_{il}\, \delta_{jk} + T^0_{lj}\, \delta_{ik}) \,,
\end{split}
\label{eq:9}
\end{equation}
which is the symmetric component of \cite{wallaceThermoelasticityStressedMaterials1967, thurstonEffectiveElasticCoefficients1965, barronSecondorderElasticConstants1965}
\begin{equation}
    \tilde C^0_{ijkl} = A^0_{ijkl} - T^0_{ij}\, \delta_{kl} + T^0_{il}\, \delta_{jk} + T^0_{lj}\, \delta_{ik} \,.
\end{equation}
For an initial state under hydrostatic stress, $T^0_{ij} = \mbox{}- P\,\delta_{ij}$, Eq.~\eqref{eq:9} reduces to \cite{barronSecondorderElasticConstants1965}
\begin{equation}
C^0_{ijkl} = A^0_{ijkl} + P \,(\delta_{ij} \,\delta_{kl} - \delta_{il}\, \delta_{kj} - \delta_{ik}\, \delta_{jl})\,,
\label{eq:10}
\end{equation}
and so does $\Gamma_{ijkl}$ in \cite{dahlenTheoreticalGlobalSeismology1998}. Therefore, different forms of the ``effective'' tensors ($B_{ijkl}$ in \cite{wallaceThermoelasticityStressedMaterials1967}, $\Upsilon_{ijkl}$ in \cite{dahlenTheoreticalGlobalSeismology1998}, and $\Gamma_{ijkl}$ in \cite{dahlenTheoreticalGlobalSeismology1998}) are equivalent under hydrostatic prestress.

\begin{widetext}
Likewise, to more conveniently evaluate the effect of strain on SOECs under hydrostatic conditions, we are motivated to introduce an effective TOE tensor.
The general effective TOE tensor $\tilde C^0_{ijklmn}$ under the infinitesimal formalism is:
\begin{equation}
\begin{split}
    \tilde C^0_{ijklmn} =&~ A^0_{ijklmn} - A^0_{ijkl}\,\delta_{mn} \\
    &+ A^0_{njkl}\,\delta_{im} + A^0_{inkl}\,\delta_{jm} + A^0_{ijnl}\,\delta_{km} + A^0_{ijkn}\,\delta_{lm} \, .
\end{split}
\label{eq:12}
\end{equation}
The elements of $\tilde C^0_{ijklmn}$ satisfy the general relationship $\Delta A^\mathrm{L}_{ijkl} = \tilde C^0_{ijklmn} (\epsilon_{mn} + \omega_{mn})$\,. Here, $F_{ij} - \delta_{ij} = \epsilon_{ij} + \omega_{ij}$\,; the antisymmetric infinitesimal tensor (or infinitesimal rotation tensor) $\omega_{ij}$ is given by $\omega_{ij} = \frac{1}{2} (F_{ij} - F_{ji})$\,.
The symmetric component of $\tilde C^0_{ijklmn}$\,,
\begin{equation}
\begin{split}
    C^0_{ijklmn} =&~ \partial A^0_{ijkl} / \partial\epsilon_{mn}
    = A^0_{ijklmn} - A^0_{ijkl}\,\delta_{mn} \\&~+ \tfrac{1}{2}(
        A^0_{njkl}\,\delta_{im} + A^0_{inkl}\,\delta_{jm} + A^0_{ijnl}\,\delta_{km} + A^0_{ijkn}\,\delta_{lm} \\
    &\quad\quad\quad
        +A^0_{mjkl}\,\delta_{in} + A^0_{imkl}\,\delta_{jn} + A^0_{ijml}\,\delta_{kn} + A^0_{ijkm}\,\delta_{ln})\, ,
\end{split}
\label{eq:12'}
\end{equation}
gives a linearized relationship with $\Delta A^\mathrm{L}_{ijkl}$ under a symmetric infinitesimal strain $\epsilon_{mn}$ that has the same form as Eq.~\eqref{eq:8}, that is,
\begin{equation}
\begin{split}
    & \Delta A_{ijkl}^\mathrm{L} \equiv A_{ijkl}^\mathrm{L} - A_{ijkl}^0 = C^0_{ijklmn} \, \epsilon_{mn} \\
    & \qquad \text{ or \quad}
    C^0_{ijklmn} = \partial A^\mathrm{L}_{ijkl} / \partial \epsilon_{mn}\,.
\end{split}
\label{eq:11}
\end{equation}
The antisymmetric component of $\tilde C^0_{ijklmn}$ characterizes the effect of rotation:
\begin{equation}
\begin{split}
    \partial A^0_{ijkl} / \partial\omega_{mn} =&~ \tfrac{1}{2}(
        A^0_{njkl}\,\delta_{im} + A^0_{inkl}\,\delta_{jm} + A^0_{ijnl}\,\delta_{km} + A^0_{ijkn}\,\delta_{lm} \\
    &\quad\quad
        -A^0_{mjkl}\,\delta_{in} - A^0_{imkl}\,\delta_{jn} - A^0_{ijml}\,\delta_{kn} - A^0_{ijkm}\,\delta_{ln})\, .
\end{split}
\label{eq:12''}
\end{equation}
\end{widetext}

In the expression for $\tilde C^0_{ijklmn}$ (Eq.~\ref{eq:12}), the contribution of $A^0_{ijklmn}$ corresponds to $\Delta A^\mathrm{SK}_{ijkl}$\,, as shown in Eqs.~\eqref{eq:7}--\eqref{eq:9-1}; for the remaining terms ($\tilde C^0_{ijklmn} - A^0_{ijklmn}$), it can be shown that (see Appendix~\ref{sec:a}): 
\begin{equation}
\begin{split}
A^\mathrm{L}_{ijkl} - A^\mathrm{SK}_{ijkl} &= J^{-1} F_{ir} F_{js} F_{kp} F_{lq} A^\mathrm{SK}_{rspq} - A^\mathrm{SK}_{ijkl} \\
&\simeq (J^{-1} F_{ir} F_{js} F_{kp} F_{lq} - \delta_{ir} \delta_{js} \delta_{kp} \delta_{lq}) \, A^0_{rspq} \\
& = (- A^0_{ijkl}\,\delta_{mn}
    + A^0_{njkl}\,\delta_{im} + A^0_{inkl}\,\delta_{jm} \\
    &\quad\quad+ A^0_{ijnl}\,\delta_{km} + A^0_{ijkn}\,\delta_{lm} ) \, (\epsilon_{mn}+\omega_{mn}) \\
& = (\tilde C^0_{ijklmn} - A^0_{ijklmn}) \, (\epsilon_{mn}+\omega_{mn}) 
    \, . 
\end{split}
\label{eq:19-1}
\end{equation}
The equality above does not impose any symmetry requirements on $A^0_{ijkl}$ other than invariance under the exchange of indices within the pairs $(i,j)$, $(k,l)$, and $(m,n)$. This invariance is guaranteed by the symmetry of the stress and strain tensors, allowing Voigt notation on these tensors \cite{huangAtomicTheoryElasticity1950, barronSecondorderElasticConstants1965}. Similar to $C^0_{ijkl}$ and $\tilde C^0_{ijkl}$, $C^0_{ijklmn}$ or $\tilde C^0_{ijklmn}$ do not have symmetries that would allow exchanges between $(i,j)$ and $(k,l)$ or $(m,n)$ pairs.

We note that the elements~$\tilde C^0_{ijklmn}$ defined in Eq.~\eqref{eq:12} are identical to the quantities~$\Theta_{ijklmn}$ in \citet[][Eq.~(A57)]{maitraStressDependenceElastic2021}. In \citet[][Appendix~A]{maitraStressDependenceElastic2021}, $\Gamma_{ijkl}$ and $\Xi_{ijkl}$ denote the thermodynamic SOEC in the initial and present frames (background and equilibrium frames in their terms); they correspond to $A^0_{ijkl}$ and $A^\mathrm{L}_{ijkl}$ in this study.

At $P = 0$~GPa, because the second term on the r.h.s.\ of Eq.~\eqref{eq:10} vanishes, we have $A^0_{ijkl} = C^0_{ijkl}$\,,
which explains why a distinction between the two types of SOECs is sometimes not made.
For TOECs, however, since $A^0_{ijkl} \neq 0$ at 0~GPa, $C^0_{ijklmn}$ and $A^0_{ijklmn}$ are never equal according to Eq.~\eqref{eq:12}. Therefore, one always needs to be specific about which TOECs are being used, even at $P = 0$~GPa. It is worth noticing that some previous reports on ``effective TOEC tensors'' \citep[e.g.,][]{krasilnikovElasticConstantsSolids2012, vekilovHigherorderElasticConstants2016, mosyaginInitioCalculationsPressuredependence2017} are available. 
Even though the ``effective SOECs'' in \cite{krasilnikovElasticConstantsSolids2012, vekilovHigherorderElasticConstants2016, mosyaginInitioCalculationsPressuredependence2017} agree with ours, their ``effective TOECs'' are not equivalent to ours. The difference between ``effective TOECs'' and ``thermodynamic TOECs'' in \cite{krasilnikovElasticConstantsSolids2012, vekilovHigherorderElasticConstants2016, mosyaginInitioCalculationsPressuredependence2017} is not the same as in the present work. This is evident when comparing these tensors at 0~GPa. The ``effective TOECs'' in \cite{krasilnikovElasticConstantsSolids2012, vekilovHigherorderElasticConstants2016, mosyaginInitioCalculationsPressuredependence2017} have the same values as the thermodynamic TOECs at 0~GPa. This is not the case for our effective TOECs ($C_{ijklmn}^0$). Recent discussions concerning these alternative forms of ``effective TOECs'' in \cite{krasilnikovElasticConstantsSolids2012, vekilovHigherorderElasticConstants2016, mosyaginInitioCalculationsPressuredependence2017} can be found in \cite{levitasNonlinearElasticityPrestressed2021, levitasReplyCommentNonlinear2022, krasilnikovCommentNonlinearElasticity2022}.


To reduce clutter, we drop the superscript $0$ in the remainder unless otherwise noted.

\section{\textit{Ab initio} calculations}
\label{sec:3}

Our \textit{ab initio} validations are performed on NaCl and MgO. Both systems are cubic and belong to the $Fm\bar{3}m$ space group. Systems within the $Fm\bar{3}m$ space group have three independent SOECs ($c_{11}$, $c_{12}$, $c_{44}$), and six independent TOECs ($c_{111}$, $c_{112}, c_{123}$, $c_{144}$, $c_{155}$, $c_{456}$) \cite{bruggerPureModesElastic1965}. Here, we calculate SOECs and TOECs for these systems under finite hydrostatic pressure.

\textit{Ab initio} calculations were performed with the Quantum ESPRESSO code suite \cite{giannozziQUANTUMESPRESSOModular2009} based on the local density approximation (LDA) \cite{perdewSelfinteractionCorrectionDensityfunctional1981} applied to the Density Functional Theory (DFT). Norm-conserving pseudopotentials generated with the Martin-Troullier method \cite{troullierEfficientPseudopotentialsPlanewave1991} were used for Na, Cl, Mg, and O. For NaCl, the energy cut-off was set to 160~Ryd and Brillouin zones were sampled with a shifted $8 \times 8 \times 8$ Monkhorst-Pack $k$-point grid; for MgO, the energy cut-off was set to 160~Ryd and Brillouin zones were sampled with a shifted $16 \times 16 \times 16$ Monkhorst-Pack $k$-point grid.

We follow the recipe of \citet{zhaoFirstprinciplesCalculationsSecond2007} to obtain the thermodynamic elastic coefficients. The expansion coefficients $M_1$, $M_2$, and $M_3$ from the polynomial expansion of energy vs.\ strain magnitude $\eta$,
\begin{equation}
\frac{E(T_{ij},\xi)}{V} = M_0 + M_1\, \eta + M_2 \,\eta^2 + M_3 \,\eta^3 + O(\eta^4) \,,
\label{eq:13}
\end{equation}
are linear combinations of $T^\mathrm{0}_{ij}$, $A_{ijkl}^\mathrm{0}$, and $A_{ijklmn}^\mathrm{0}$ for the configuration under initial stress $T_{ij}$. The $M_1\, \eta$ terms are necessary to account for the initial stress in this study (in this instance, the hydrostatic prestress $T^0_{ij} = \mbox{}-P\,\delta_{ij}$\,); they equal zero for the 0~GPa state and are thus absent in earlier studies \cite{zhaoFirstprinciplesCalculationsSecond2007}.
Perturbations with a set of linearly-independent Lagrangian strains $\eta_{ij}$ of the form (A1--A6) \cite{zhaoFirstprinciplesCalculationsSecond2007} with magnitudes $\eta = 0.00, \pm 0.01, \pm 0.02, \pm 0.03$ are applied. The corresponding symmetric part of the deformation gradient $F_{ij}$ is obtained from the Lagrangian strain $\eta_{ij}$ using a scheme described by Ref.~\cite{liaoElastic3rdToolCalculating2021}.
$T^0_{ij}$, $A^0_{ijkl}$, and $A^0_{ijklmn}$ are determined by inverting the linear equations in Table~I of Ref.~\cite{zhaoFirstprinciplesCalculationsSecond2007}.

Fig.~S1 compares $\mbox{}-T^0_{11}$ vs.\ $V$ from the strain energy expansion in $P$ of the third-order Birch-Murnaghan finite-strain equation of state fitted given $E$ vs.\ $V$ (i.e., $P = \mbox{}-\partial E / \partial V$). The hydrostatic condition $\mbox{}-T^0_{11} = P$ is observed here by the excellent consistency between the two.

\begin{figure*}[htbp]
\centering

\begin{tabular}{@{}l@{}}
\includegraphics[width=.4\textwidth]{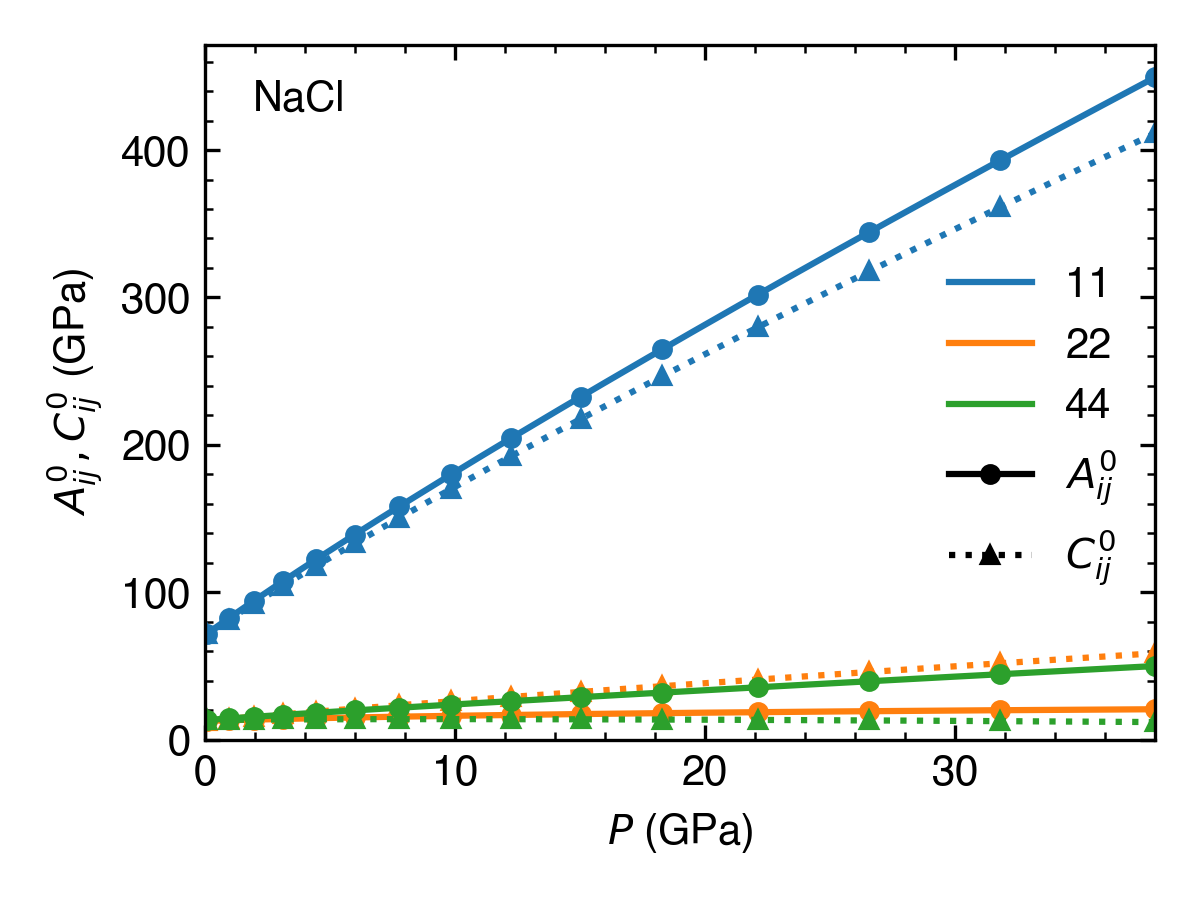}
\includegraphics[width=.4\textwidth]{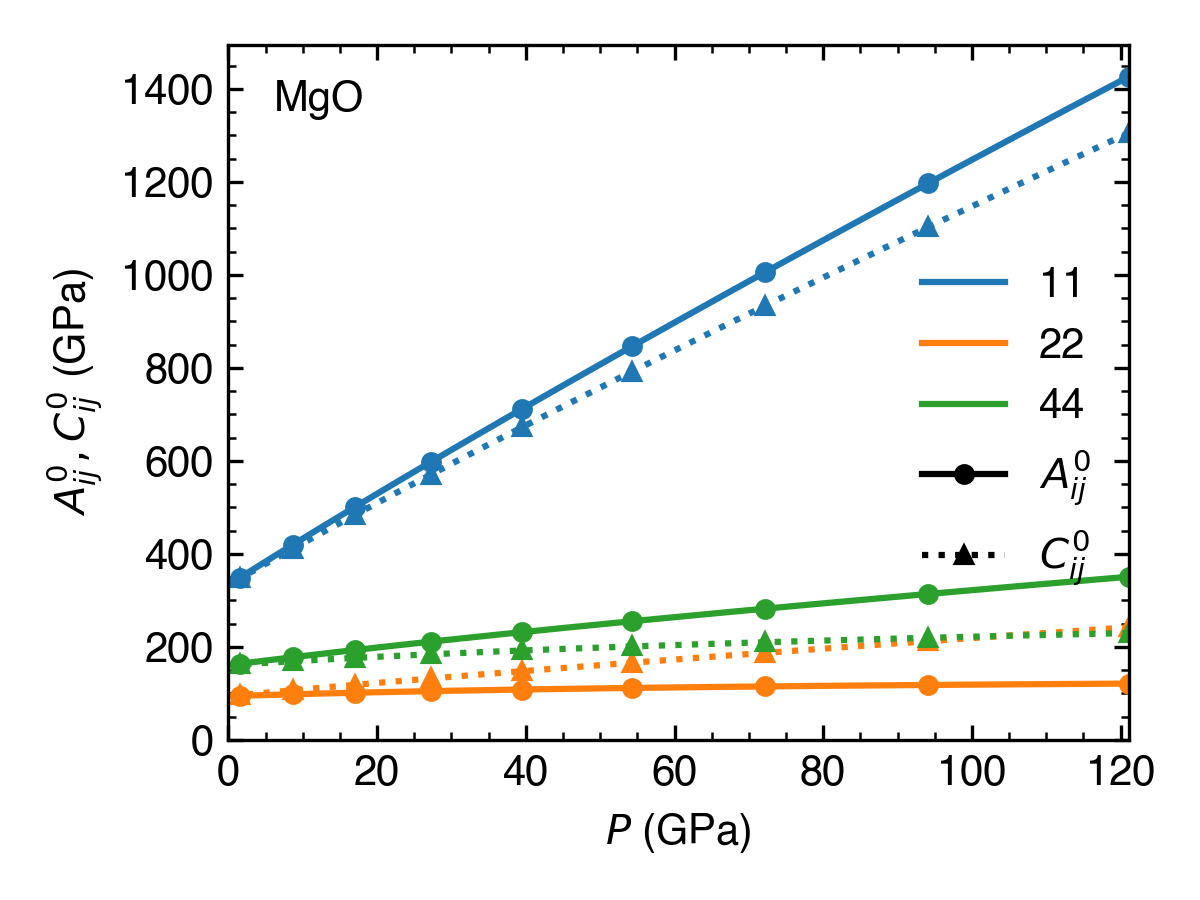} \\

\includegraphics[width=.4\textwidth]{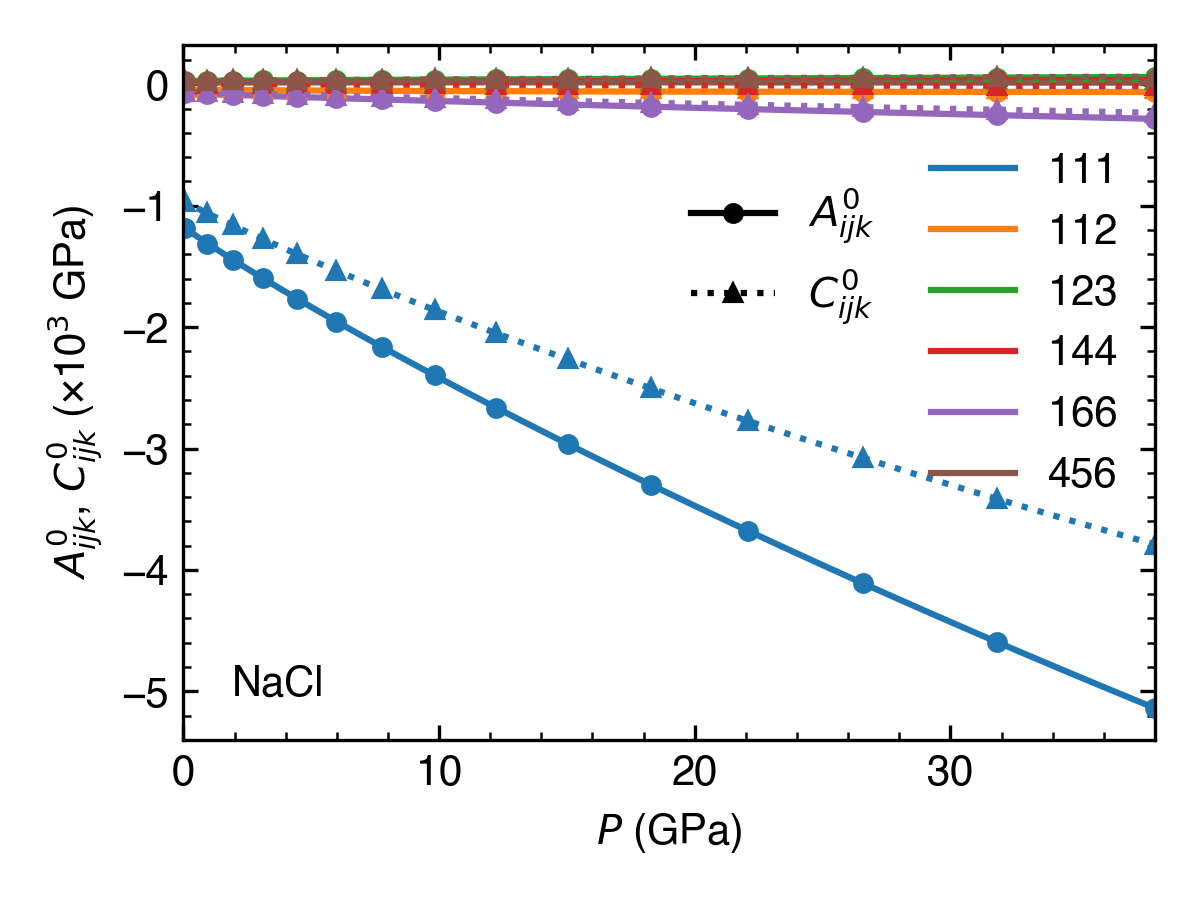}
\includegraphics[width=.4\textwidth]{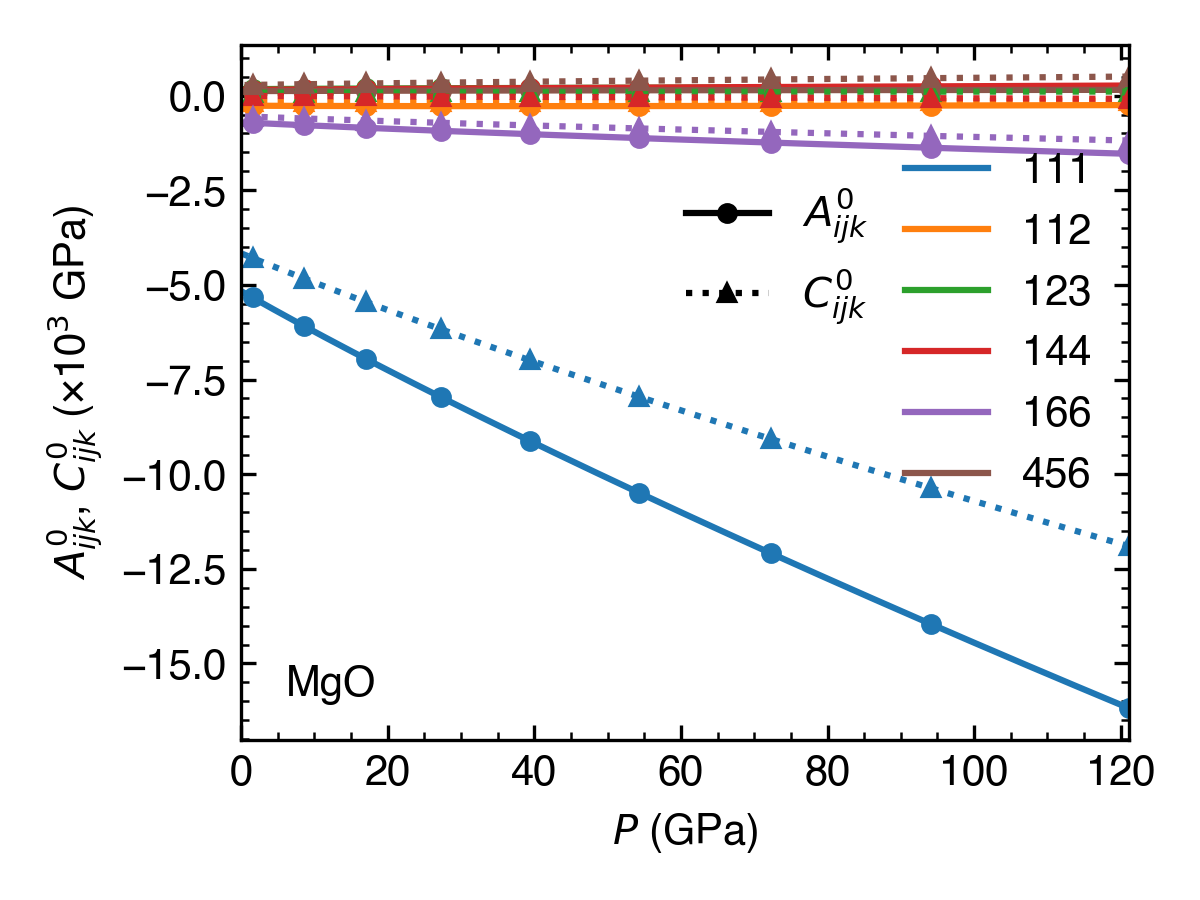}
\end{tabular}

\caption{(a, b) SOECs and (c, d) TOECs vs.\ $P$ for (a, c) NaCl and (b, d) MgO. Tensor indices are in Voigt notation.}
\label{fig:1}
\end{figure*}

Fig.~\ref{fig:1} shows the calculated SOECs (a, b) and TOECs (c, d) vs.\ $P$\,. The thermodynamic elastic coefficients are obtained directly from the energy vs.\ strain expansion (Eq.~\eqref{eq:13}); the effective elastic coefficients are calculated based on Eqns.~\eqref{eq:9}, \eqref{eq:10}, and \eqref{eq:12}. The calculated data points for SOECs and TOECs are displayed as scattered symbols. Overall, the expansion of energy vs.\ strain is a robust and effective approach to computing elastic coefficients for cubic systems up to third-order at finite pressure. Interpolated elastic coefficients, shown as smooth curves, are used to determine values at intermediate volumes or pressures.
The TOECs have a greater magnitude than the SOECs. For NaCl and MgO, the TOECs and SOECs are a near-linear function of pressure.

\section{Effect of strain on SOECs}
\label{sec:4}

In this section,
we evaluate the effect of strain on thermodynamic SOECs based on the \textit{ab initio} calculated TOECs discussed in the previous section, using Eq.~\eqref{eq:7} to obtain $\Delta A^\mathrm{SK}$ and \eqref{eq:11} for $\Delta A^\mathrm{L}$\,.
We are not going to address effects on effective SOECs, but the additional stress related-terms in Eq.~\eqref{eq:9} can be easily calculated by substituting the induced stress calculated from Eq.~\eqref{eq:8} into Eq.~\eqref{eq:9}.

Previously, Refs.~\cite{trompEffectsInducedStress2018, trompEffectsInducedStress2019} have shown that if the induced stress $T^0_{ij}$ is known,
$\Delta A^\mathrm{L}_{ijkl}$ and $\Delta A^\mathrm{SK}_{ijkl}$
can be evaluated based on the pressure derivative of the thermodynamic SOECs, $A^{0\,\prime}_{ijkl} \equiv \partial A^0_{ijkl} / \partial P$\,, as
\begin{equation}
\begin{split}
    \Delta A^\mathrm{L}_{ijkl} =&~ A_{ijkl}^{0\,\prime} \, p - \tfrac{1}{4}(A^{0\,\prime}_{mjkl}\, \tau_{im} + A^{0\,\prime}_{imkl}\, \tau_{jm} \\&\quad+ A^{0\,\prime}_{ijml}\, \tau_{km} + A^{0\,\prime}_{ijkm}\, \tau_{lm}) \,,
\end{split}
\label{eq:14}
\end{equation}
where the hydrostatic stress is given by~$p = \mbox{}-\frac{1}{3}\, \mathrm{tr}(\mathbf{T}^0)$, and the deviatoric stresses are given by~$\tau_{ij} = T^0_{ij} + p \,\delta_{ij} = T^0_{ij} - \tfrac{1}{3}\, \mathrm{tr}(\mathbf{T}^0)\, \delta_{ij}$\,. Since the pre-stress $T^\mathrm{L}_{ij}$ under $\epsilon_{mn}$ can be evaluated based on Eq.~\eqref{eq:8}, this method offers a viable alternative to evaluating $\Delta A^\mathrm{L}_{ijkl}$ vs.\ $\epsilon_{mn}$.
For comparison,
this scheme will also be included in our validations.

We consider the practical situation where the initial configuration is under hydrostatic pressure. In this scenario, the SOECs and TOECs for the initial configuration are already known and have familiar cubic symmetry \cite{bruggerPureModesElastic1965, barronSecondorderElasticConstants1965}.
Changes in SOECs induced by strain in two different forms are addressed: (a) under $\epsilon_{11} = \epsilon_{22} = \epsilon_{33} = 0.005$, that is, a uniform stretch, and (b) under $\epsilon_{33} = 0.005$, $\epsilon_{11} = \epsilon_{22}=0$, that is, a uniaxial stretch.

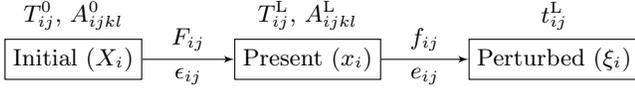
\begin{figure}[h]
    \centering
  

\begin{tikzpicture}
\tikzstyle{every node}=[font=\small]
\node (s0) [draw, label = {$T^0_{ij}$, $A^0_{ijkl}$} ] {Initial ($X_i$)};
\node (s1) [draw, label = {$T^\mathrm{L}_{ij}$, $A^\mathrm{L}_{ijkl}$}, right of=s0, right=3.5em] {Present ($x_i$)};
\node (s2) [draw, label = {$t^\mathrm{L}_{ij}$}, right of=s1, right=3.5em] {Perturbed ($\xi_i$)};
\path [line] (s0) -- node [anchor=south] {$F_{ij}$} node [anchor=north] {$\epsilon_{ij}$} (s1);
\path [line] (s1) -- node [anchor=south] {$f_{ij}$} node [anchor=north] {$e_{ij}$} (s2);
\end{tikzpicture}
    
\caption{Relevant configurations for validating the SOECs vs.\ strain relationships.
\textsl{Left\/}: The initial configuration with coordinates \{$X_i$\} is in equilibrium with the initial hydrostatic pressure; the related SOECs and TOECs are known. We adopt the hydrostatically stressed configuration discussed in Section~\ref{sec:3} as the initial configuration.
\textsl{Middle\/}: The present configuration with coordinates \{$x_i$\} is derived from the initial configuration \{$X_i$\} by applying an elastic deformation gradient $F_{ij} = \partial x_i / \partial X_j$, or uniform or uniaxial stretches, $\epsilon_{ij}$; they are the ones whose SOECs are in question; the external stress $T^\mathrm{L}_{ij}$ in equilibrium with this configuration can be calculated from the stress vs.\ strain relationship.
\textsl{Right\/}: The perturbed \{$\xi_i$\} configuration is invoked when necessary; this is achieved by perturbing the \{$x_i$\} configuration with the elastic deformation gradient $f_{ij} = \partial \xi_i / \partial x_j $\,,
or infinitesimal strain~$e_{ij}$\,.
}
\label{fig:5}
\end{figure}

Fig.~\ref{fig:5} summarizes the three relevant configurations for the validation.


The specifics for calculating SOECs in the present configuration for the two forms of strains tested and comparisons between the TOE-predicted and numerically-evaluated elastic tensors $\Delta A^\mathrm{SK}_{ijkl}$ and $\Delta A^\mathrm{L}_{ijkl}$ are as follows.

\paragraph{Uniform stretch}

For a cubic system under hydrostatic pressure, the effect of a uniform stretch corresponds to decreasing the external pressure.
Therefore, the elastic coefficients for such configurations are already available from our previous interpolation of the elastic coefficients vs.\ volume. For an initial configuration with volume $V_0$, the corresponding present volume $V$ under the stretch $\epsilon_{11} = \epsilon_{22} = \epsilon_{33} = \epsilon = 0.005$ is $V = (1 + \epsilon)^3 \, V_0 \approx (1+3\,\epsilon)\, V_0$.

\begin{figure*}[htbp]
\centering

\input{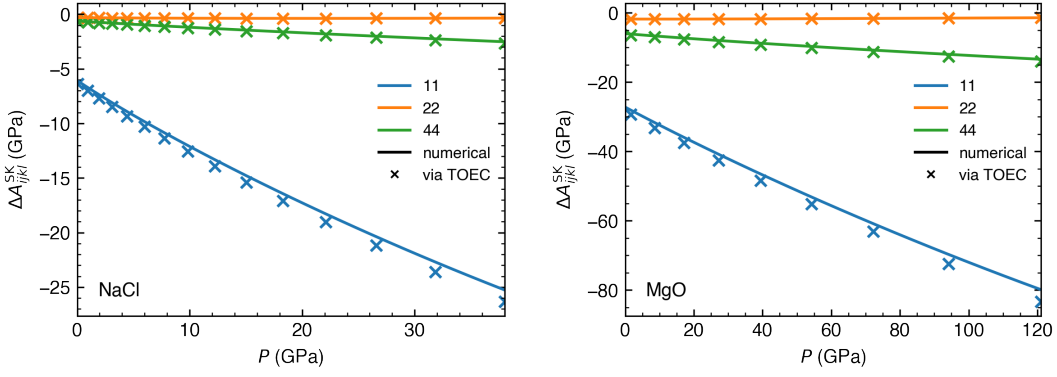}
\caption{Change in elastic coefficients $\Delta A^\mathrm{SK}_{ijkl}$ induced by a uniform stretch, $\epsilon_{11} = \epsilon_{22} = \epsilon_{33} = 0.005$.
Tensor indices are in Voigt notation.}
\label{fig:3}
\end{figure*}

\begin{figure*}[htbp]
\centering
\input{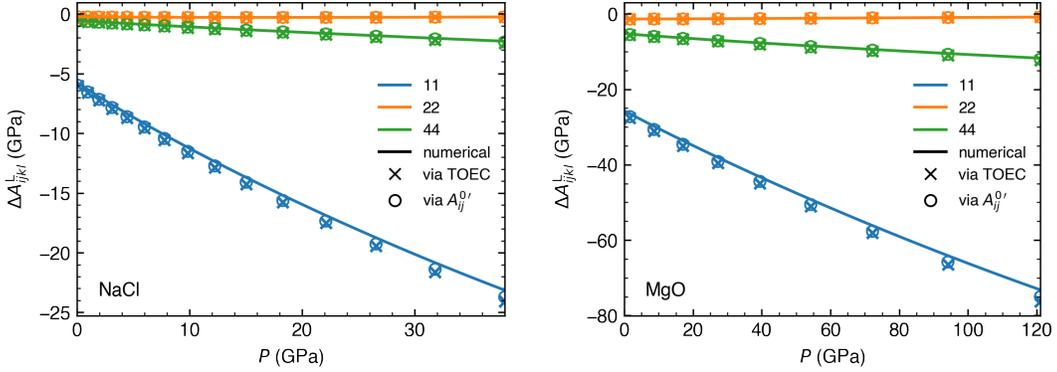}
\caption{Change in elastic coefficients $\Delta A^\mathrm{L}_{ijkl}$ components induced by a uniform stretch, $\epsilon_{11} = \epsilon_{22} = \epsilon_{33} = 0.005$.
Tensor indices are in Voigt notation.}
\label{fig:4}
\end{figure*}

Fig.~\ref{fig:3} shows $\Delta A^\mathrm{SK}_{ijkl}$ vs.\ $P$, and 
Fig.~\ref{fig:4} shows $\Delta A^\mathrm{L}_{ijkl}$ vs.\ $P$.
The prediction of $\Delta A^\mathrm{SK}_{ijkl}$ based on $A^0_{ijklmn}$\,, according to Eq.~\eqref{eq:7}, and the prediction of $\Delta A^\mathrm{L}_{ijkl}$ based on  $C^0_{ijklmn}$\,,  according to Eq.~\eqref{eq:11}, are both in good agreement with the numerically calculated result. Whether evaluated within a uniform reference, the prediction based on the TOEC tensors shows comparable accuracy, provided the correct set of tensors is used.

\paragraph{Uniaxial stretch}

Under uniaxial stretch, the present configuration no longer has the $m\bar{3}m$ symmetry, we need to compute the SOECs for the present configuration ($A^\mathrm{L}_{ijkl}$) first. This can be achived by using the second Piola-Kirchhoff stress $t^\mathrm{SK1}_{ij}$ vs.\ strain $e_{kl}$ relation, whose expression is similar to Eq.~\eqref{eq:7}:
\[
t^\mathrm{SK1}_{ij} = A^\mathrm{L}_{ijkl} \, e_{kl}\,.
\]
A total of 6 sets of perturbed configurations (\{$\xi_i$\}) is used to obtain the full $A^\mathrm{L}_{ijkl}$ tensor.
Measured within the present frame, the incremental second Piola-Kirchhoff stress for the perturbed configuration $t^\mathrm{SK1}_{ij}$ is given by \[ t^\mathrm{SK1}_{ij} = t^\mathrm{SK}_{ij} - t^0_{ij} = t^\mathrm{SK}_{ij} - T^\mathrm{L}_{ij}\,, \]
because the present configuration's Lagrangian Cauchy stress is the initial stress within the present reference frame,
that is, $t^0_{ij} = T^\mathrm{L}_{ij}$\,.
Similar to Eq.~\eqref{eq:6}, the expression for pulling back $t^\mathrm{L}_{ij}$ from the perturbed frame to $t^\mathrm{SK}_{ij}$ within the present frame is
\[  t^\mathrm{SK}_{ij} = j \, f^{-1}_{ir} f^{-1}_{js}\, t^\mathrm{L}_{rs} = j \, \frac{\partial x_i}{\partial \xi_{r}} \frac{\partial x_j}{\partial \xi_{s}} \,t^\mathrm{L}_{rs} \,.  \]
where $f_{ir} = \partial \xi_r / \partial x_i$,
and the Jacobian $j = \det \mathbf{f}$.

\begin{figure*}[htbp]
\centering
\input{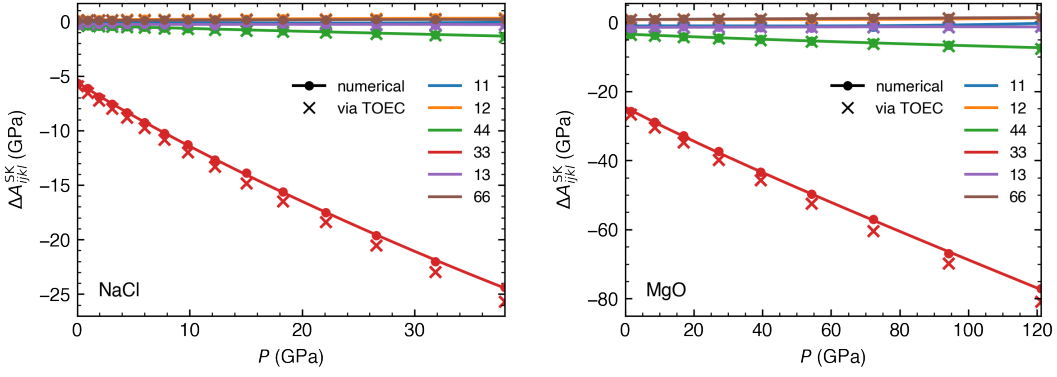}
\caption{Change in elastic coefficients $\Delta A^\mathrm{SK}_{ijkl}$ induced by a uniaxial stretch, $\epsilon_{11} = \epsilon_{22}=0$, $\epsilon_{33}=0.005$.
Tensor indices are in Voigt notation.}
\label{fig:6}
\end{figure*}

\begin{figure*}[htbp]
\centering
\input{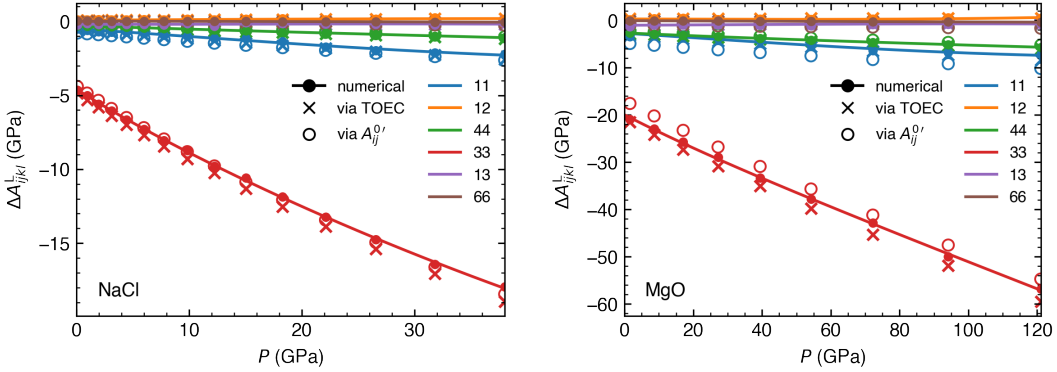}
\caption{Change in elastic coefficients $\Delta A^\mathrm{L}_{ijkl}$ induced by a uniaxial stretch, $\epsilon_{11} = \epsilon_{22} = 0$, $\epsilon_{33}=0.005$.
 Tensor indices are in Voigt notation.}
\label{fig:7}
\end{figure*}

With the SOECs $A^\mathrm{L}_{ijkl}$ and $A^\mathrm{0}_{ijkl}$ both available via numerical calculations, we compare their difference vs.\ pressure with TOE predictions.
Fig.~\ref{fig:6} shows $\Delta A^\mathrm{SK}_{ijkl}$ vs.\ pressure computed in a uniform reference frame. Fig.~\ref{fig:7} shows $\Delta A^\mathrm{L}_{ijkl}$ vs.\ pressure with elastic coefficients before and after the strain computed in their own frames. Overall, our predictions based on Eq.~\eqref{eq:16} are in good agreement with numerically calculated values. Breaking of the cubic ($m\bar{3}m$) symmetry results in the splitting of $A_{33}$ from $A_{11}$, $A_{13}$ from $A_{12}$, and $A_{66}$ from $A_{44}$.

Under uniform and uniaxial stretches, TOE theory accurately predicts the incremental SOECs for configurations not far from the initial condition. Residuals in both $\Delta A^\mathrm{SK}_{ijkl}$ and $\Delta A^\mathrm{L}_{ijkl}$ originate from approximating changes in SOECs linearly with TOECs $A^0_{ijklmn}$ and $C^0_{ijklmn}$, both of which are also functions of strain.


\section{Pressure derivatives of SOECs}
\label{sec:5}

TOE theory allows us to assess the pressure derivatives of SOECs. We discuss pressure derivatives of the thermodynamic SOECs, $A^{0\,\prime}_{ijkl}$\,, only. Pressure derivatives of the effective SOECs $C^{0\,\prime}_{ijkl}$ may be obtained by addition of the terms $(\delta_{ij}\, \delta_{kl} - \delta_{il} \,\delta_{kj} - \delta_{ik} \,\delta_{jl})$\,. 

First, we derive an expression for the pressure derivatives $A^{0\,\prime}_{ijkl}$\, based on the TOECs $A_{ijklmn}^\mathrm{0}$\,.
Noting that on the right-hand side of Eq.~\eqref{eq:5-2}, $E$ and $V$ are both functions of $P$, we have
\begin{equation}
\begin{split}
A^{0\,\prime}_{ijkl} &=\frac{\partial A_{ijkl}^\mathrm{0}}{\partial P}
= \frac{\partial}{\partial P}\left(\frac{1}{V} \frac{\partial^{2} E}{\partial \eta_{ij} \partial \eta_{kl}}\right) \\
&= \frac{\partial}{\partial P}\left(\frac{1}{V}\right) \frac{\partial^{2} E}{\partial \eta_{ij} \partial \eta_{kl}} + \frac{1}{V} \frac{\partial}{\partial P}\left(\frac{\partial^{2} E}{\partial \eta_{ij} \partial \eta_{kl}}\right)\\
&= \mbox{}-\frac{1}{V^{2}} \frac{\partial V}{\partial P}\,
\frac{\partial^{2} E}{\partial \eta_{ij} \partial \eta_{kl}}
+\frac{1}{V}\left(\frac{\partial^3 E}{\partial\eta_{ij} \partial\eta_{kl} \partial\eta_{mn}}\right) \frac{\partial \eta_{mn}}{\partial P}
\,.
\end{split}
\label{eq:16}
\end{equation}
Because the initial and final states are both under hydrostatic conditions ($T_{ij} = \mbox{}-P\,\delta_{ij}$), the stress vs.\ strain relation Eq.~\eqref{eq:7} determines changes in stress (or pressure) caused by a strain $\Delta \eta_{mn}$ via
$C_{mnop}^\mathrm{0}\, \Delta \eta_{op} = \mbox{}-\Delta P\, \delta_{mn}$\,.
Using the compliance tensor $(C^{-1})_{mnop}$\,, we have
$\Delta \eta_{mn} = \mbox{}- (C^{-1})_{mnop}\, \Delta P \,\delta_{op}$\,,
and therefore,
\begin{equation}
\frac{\partial \eta_{mn}}{\partial P} = \mbox{}-\left(C^{-1}\right)_{mnop}\, \delta_{op}\, ,
\end{equation}
\begin{equation}
\frac{1}{V} \frac{\partial V}{\partial P} =  \tfrac{1}{3}\, \frac{\partial \eta_{mn}}{\partial P}\, \delta_{mn} =  \mbox{}- \tfrac{1}{3} \left(C^{-1}\right)_{mnop}\, \delta_{mn}\, \delta_{op}\, .
\end{equation}
Thus, Eq.~\eqref{eq:10} may be simplified to become
\begin{equation}
A^{0\,\prime}_{ijkl} = \mbox{}- A_{ijklmn}^\mathrm{0}\, (C^{-1})_{mnpq}\, \delta_{pq} - \tfrac{1}{3}\, A_{ijkl}^\mathrm{0} \,(C^{-1})_{mnpq}\, \delta_{mn}\, \delta_{pq} \,,
\label{eq:21}
\end{equation}
in agreement with previously reported results \cite{barschAdiabaticIsothermalIntermediate1967, changNonlinearPressureDependence1967}.

Alternatively, we can derive a $C_{ijklmn}^\mathrm{0}$-based expression for $A^{0\,\prime}_{ijkl}$\,, because
\begin{equation}
\begin{split}
        A^{0\,\prime}_{ijkl} \ = & \  \frac{\partial A_{ijkl}^\mathrm{0}}{\partial \eta_{mn}} \,\frac{\partial \eta_{mn}}{\partial P} \,,
    \\ = & \  \mbox{}- C_{ijklmn}^\mathrm{0}\, (C^{-1})_{mnpq}\, \delta_{pq}\, .
\end{split}
\label{eq:25}
\end{equation}
This expression is identical to \citet[][Eq.~(A59)]{maitraStressDependenceElastic2021} under hydrostatic prestress.

\begin{figure*}[htbp]
\centering
\input{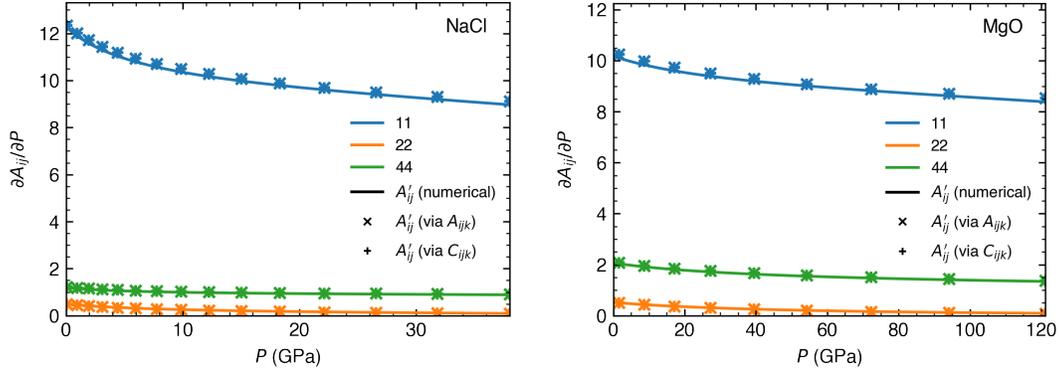}
\caption{Pressure derivative of SOECs vs.\ pressure for NaCl and MgO. Curves denote numerical pressure derivatives of the elastic coefficients; $\times$ indicates predictions based on Eq.~\eqref{eq:21};
$+$ indicates predictions based on Eq.~\eqref{eq:25}
.}
\label{fig:2}
\end{figure*}

Fig.~\ref{fig:2} shows $A^{0\,\prime}_{ijkl}$ calculated three ways. 1) Numerically calculated derivatives based on the interpolated SOECs $A_{ijkl}$ (solid curves). 2) Predictions with $A_{ijklmn}^\mathrm{0}$ based on Eq.~\eqref{eq:21} shown as ``$\times$''. 3) Predictions with $C_{ijklmn}^\mathrm{0}$ based on Eq.~\eqref{eq:25} shown as ``$+$''. Good consistency between the three methods along the entire pressure range indicates that Eqns.~\eqref{eq:21} and~\eqref{eq:25} are accurate predictions of $A^{0\,\prime}_{ijkl}$ for MgO and NaCl.

The above validation shows that $A^{0\,\prime}_{ijkl}$ at finite pressure is a linear combination of TOECs, which is why its inverse, i.e., Eq.~\eqref{eq:14}, works.

Finally, although Eq.~\eqref{eq:25} has a similar form as~\citet[][Eq.~(71)]{thurstonCalculationLatticeParameter1967}, they do not have the same meaning. In \citep{thurstonCalculationLatticeParameter1967, truesdellMechanicsSolidsVolume1984}, all SOECs are measured within the natural frame based on the second Piola-Kirchhoff description. Thus, their pressure derivatives are not the $A^{0\,\prime}_{ijkl}$ defined here, but rather 
$\lim_{P \to 0} \Delta A^\mathrm{SK}_{ijkl}\,/P$\,; their TOECs are at 0~GPa, so their predictions remain valid only within close vicinity of zero pressure.

\section{Conclusion}
\label{sec:conclusion}

In this study, we examine third-order elasticity (TOE) theory to evaluate the effects of elastic deformation on second-order elastic coefficients (SOECs). We review definitions of thermodynamic SOECs, thermodynamic TOECs, and effective SOECs under finite pressure. Based on effective SOECs, we propose the use of effective TOECs. Explicit expressions for the effective TOECs are given and verified. We extend the method to compute TOECs under finite pressure via \textit{ab initio} calculations. Based on \textit{ab initio}-calculated TOECs, we predict the effects of strain on SOECs and the pressure derivative of SOECs for two cubic systems, NaCl and MgO. Our results show that both thermodynamic TOECs and effective TOECs accurately predict strain-induced changes in SOECs. Our study also serves as a self-consistent validation of the \textit{ab initio} approach for computing TOECs.

\section*{Acknowledgments}

This research was supported by DOE award DE-SC0019759 and NSF award EAR-2000850.
This work used the Extreme Science and Engineering Discovery Environment (XSEDE) \cite{townsXSEDEAcceleratingScientific2014} \textit{Expanse} supercomputer  at the San Diego Supercomputing Center and the \textit{Bridges-2} supercomputer at the Pittsburgh Supercomputing Center through allocation TG-DMR180081.

\begin{sidewaystable*}
    \renewcommand{\arraystretch}{1.1}
    \caption{Comparison of notations used in different studies.}
    \label{tab:1}
    \centering
    \begin{ruledtabular}
\begin{tabular}{rccccccccc}
    &
    This study &
    Ref.~\cite{dahlenTheoreticalGlobalSeismology1998} &
    Ref.~\cite{birchFiniteElasticStrain1947} &
    Ref.~\cite{barronSecondorderElasticConstants1965} &
    Refs.~\cite{thurstonEffectiveElasticCoefficients1965, truesdellMechanicsSolidsVolume1984} &
    Ref.~\cite{zhaoFirstprinciplesCalculationsSecond2007} &
    Ref.~\cite{barschAdiabaticIsothermalIntermediate1967} &
    Ref.~\cite{wentzcovitchFirstPrinciplesQuasiharmonic2010} &
    Ref.~\cite{wallaceThermoelasticityStressedMaterials1967} \\
    \hline
    
    Initial frame &
    $X_i$ &
    &
    $a_s$ &
    &
    $X_i$ &
    &
    $x_i$ &
    $x_i$ &
    $x_i$ \\
    
    Present frame &
    $x_i$ &
    &
    $x_r$ &
    &
    $x_i$ &
    &
    $a_i$ &
    $X_i$ &
    $X_i$ \\
    
    \hline
    
    Deformation matrix &
    $F$ &
    $\mathbf{F}$ &
    $\partial x_r / \partial a_s$ &
    $u_{\alpha,\beta}$ &
    $\partial x_s / \partial X_i$ &
    $F_{ij}$ &
    &
    $u_{ij}$ &
    $u_{ij}$ \\

    Lagrangian strain &
    $\eta_{ij}$ &
    $\mathbf{E}^\mathrm{L}$ &
    $\eta_{rs}$ &
    $\eta_{\alpha\beta}$ &
    $S_{ij}$ \cite{thurstonEffectiveElasticCoefficients1965} or $V_{ij}$ \cite{truesdellMechanicsSolidsVolume1984} &
    $\eta_{ij}$ &
    $\eta_{ij}$ &
    $e_{ij}$ &
    $\eta_{ij}$ \\
    \hline

    Thermodynamic SOEC &
    $A_{ijkl}^\mathrm{0}$ &
    $\Xi_{ijkl}$ &
    $c_{ij}$ &
    $\mathring{C}_{\alpha\beta\sigma\tau}$ &
    $\bar{C}_{ijkm}$ &
    $C_{ij}$ / $C_{ijkl}$ &
    $C_{ijkl} - p D_{ijkl}$* &
    &
    $C_{ijkl}$ \\
    
    Effective SOEC &
    $C_{ijkl}^\mathrm{0}$ &
    $\Upsilon_{ijkl}$ &
    &
    $\mathring{c}_{\alpha\beta\sigma\tau}$ &
    $\beta_{ijkm}$ &
    &
    $C_{ijkl}$ &
    $C_{ijkl}$ &
    $B_{ijkl}$ \\
    
    Thermodynamic TOEC &
    $A_{ijklmn}^\mathrm{0}$ &
    &
    $C_{ijk}$ &
    &
    &
    $C_{ijk}$ / $C_{ijklmn}$
    &
    $C_{ijklmn}$ &
    &
    $C_{ijkl}$ \\
    
    Effective TOEC &
    $C_{ijklmn}^\mathrm{0}$ &
    &
    &
    &
    &
    &
    \\
    
    \end{tabular}
    \end{ruledtabular}    
\raggedright
* $D_{ijkl} = (\delta_{ij}\, \delta_{kl} - \delta_{il}\, \delta_{kj} - \delta_{ik}\, \delta_{jl})$
\end{sidewaystable*}

\appendix

\begin{widetext}
\section{Verification of Eq.~\eqref{eq:19-1}}

\label{sec:a}

To verify the equality in the following equation:
\begin{equation}
\begin{split}
& (J^{-1} F_{ir} F_{js} F_{kp} F_{lq} - \delta_{ir} \delta_{js} \delta_{kp} \delta_{lq}) \, A^0_{rspq} \\
& \quad\quad = \large( - A^0_{ijkl}\,\delta_{mn} +
        A^0_{njkl}\,\delta_{im} + A^0_{inkl}\,\delta_{jm} \\
        &\quad\quad\quad + A^0_{ijnl}\,\delta_{km} + A^0_{ijkn}\,\delta_{lm}
        ) \, (\epsilon_{mn} + \omega_{mn}) \,,
\end{split}
\label{eq:a1}
\end{equation}
these basic cases are studied.
\begin{enumerate}

    \item Under a hydrostatic stretch,
\[
F_{ij} = \begin{bmatrix}
e & 0 & 0 \\
0 & e & 0 \\
0 & 0 & e \\
\end{bmatrix}_{ij} + \delta_{ij} \quad (e \ll 1) \, .
\]
in Voigt notation, both sides of Eq.~\eqref{eq:a1} are
\[
A^0_{ijkl} \, e \, .
\]

    \item Under a uniaxial stretch,
\[
F_{ij} = \begin{bmatrix}
e & 0 & 0 \\
0 & 0 & 0 \\
0 & 0 & 0 \\
\end{bmatrix}_{ij} + \delta_{ij} \quad (e \ll 1) \, ,
\]
in Voigt notation, both sides of Eq.~\eqref{eq:a1} are
\[
\begin{bmatrix}
3A^0_{11} & A^0_{12} & A^0_{13} & A^0_{14} & 2A^0_{15} & 2A^0_{16} \\
A^0_{21} & -A^0_{22} & -A^0_{23} & -A^0_{24} & 0 & 0 \\
A^0_{31} & -A^0_{32} & -A^0_{33} & -A^0_{34} & 0 & 0 \\
A^0_{41} & -A^0_{42} & -A^0_{43} & -A^0_{44} & 0 & 0 \\
2A^0_{51} & 0 & 0 & 0 & A^0_{55} & A^0_{65} \\
2A^0_{61} & 0 & 0 & 0 & A^0_{65} & A^0_{66} \\
\end{bmatrix} \, e \, .
\]

\item Under a shear deformation
\[
F_{ij} = \begin{bmatrix}
0 & e & 0 \\
e & 0 & 0 \\
0 & 0 & 0 \\
\end{bmatrix}_{ij} + \delta_{ij} \quad (e \ll 1) \, ,
\]
in Voigt notation, both sides of Eq.~\eqref{eq:a1} are
\[
\tiny
\begin{bmatrix}
2A^0_{16} + 2A^0_{61} & 2A^0_{16} + 2A^0_{62} & 2A^0_{63} & A^0_{15} + 2A^0_{64} & A^0_{14} + 2A^0_{65} & A^0_{11} + A^0_{12} + 2A^0_{66} \\
2A^0_{26} + 2A^0_{61} & 2A^0_{26} + 2A^0_{62} & 2A^0_{63} & A^0_{25} + 2A^0_{64} & A^0_{24} + 2A^0_{65} & A^0_{21} + A^0_{22} + 2A^0_{66} \\
2A^0_{36} & 2A^0_{36} & 0 & A^0_{35} & A^0_{34} & A^0_{31} + A^0_{32} \\
2A^0_{46} + A^0_{51} & 2A^0_{46} + A^0_{52} & A^0_{53} & A^0_{45} + A^0_{54} & A^0_{44} + A^0_{55} & A^0_{41} + A^0_{42} + A^0_{56} \\
A^0_{41} + 2A^0_{56} & A^0_{42} + 2 A^0_{56} & A^0_{43} & A^0_{44} + A^0_{55} & A^0_{45} + A^0_{54} & A^0_{46} + A^0_{51} + A^0_{52} \\
A^0_{11} + A^0_{21} + 2A^0_{66} & A^0_{12} + A^0_{22} + A^0_{66} & A^0_{13} + A^0_{23} & A^0_{14} + A^0_{24} + A^0_{65} & A^0_{15} + A^0_{25} + A^0_{64} & A^0_{16} + A^0_{26} + A^0_{61} + A^0_{62}
\end{bmatrix}\,e \, .
\]

\item Under a rotation
\[
F_{ij} = \begin{bmatrix}
0 & e & 0 \\
-e & 0 & 0 \\
0 & 0 & 0 \\
\end{bmatrix}_{ij} + \delta_{ij} \quad (e \ll 1) \, ,
\]
in Voigt notation, both sides of Eq.~\eqref{eq:a1} are
\[
\tiny
\begin{bmatrix}2 A_{16} + 2 A_{61} & - 2 A_{16} + 2 A_{62} & 2 A_{63} & -  A_{15} + 2 A_{64} &  A_{14} + 2 A_{65} & -  A_{11} +  A_{12} + 2 A_{66}\\
2 A_{26} - 2 A_{61} & - 2 A_{26} - 2 A_{62} & - 2 A_{63} & -  A_{25} - 2 A_{64} &  A_{24} - 2 A_{65} & -  A_{21} +  A_{22} - 2 A_{66}\\
2 A_{36} & - 2 A_{36} & 0 & -  A_{35} &  A_{34} & -  A_{31} +  A_{32}\\
2 A_{46} -  A_{51} & - 2 A_{46} -  A_{52} & -  A_{53} & -  A_{45} -  A_{54} &  A_{44} -  A_{55} & -  A_{41} +  A_{42} -  A_{56}\\
A_{41} + 2 A_{56} &  A_{42} - 2 A_{56} &  A_{43} &  A_{44} -  A_{55} &  A_{45} +  A_{54} &  A_{46} -  A_{51} +  A_{52}\\
-  A_{11} +  A_{21} + 2 A_{66} & -  A_{12} +  A_{22} - 2 A_{66} & -  A_{13} +  A_{23} & -  A_{14} +  A_{24} -  A_{65} & -  A_{15} +  A_{25} +  A_{64} & -  A_{16} +  A_{26} -  A_{61} +  A_{62}
\end{bmatrix}e \, .
\]

\end{enumerate}
\end{widetext}

\bibliography{Geophysics}

\normalcolor


\end{document}


\normalcolor
\widetext

\setcounter{section}{0}
\renewcommand{\thesection}{S-\Roman{section}}
\setcounter{figure}{0}
\renewcommand{\thefigure}{S\arabic{figure}}

\section*{Supplementary materials}

\begin{figure*}[h]
\centering
\includegraphics[width=.4\textwidth]{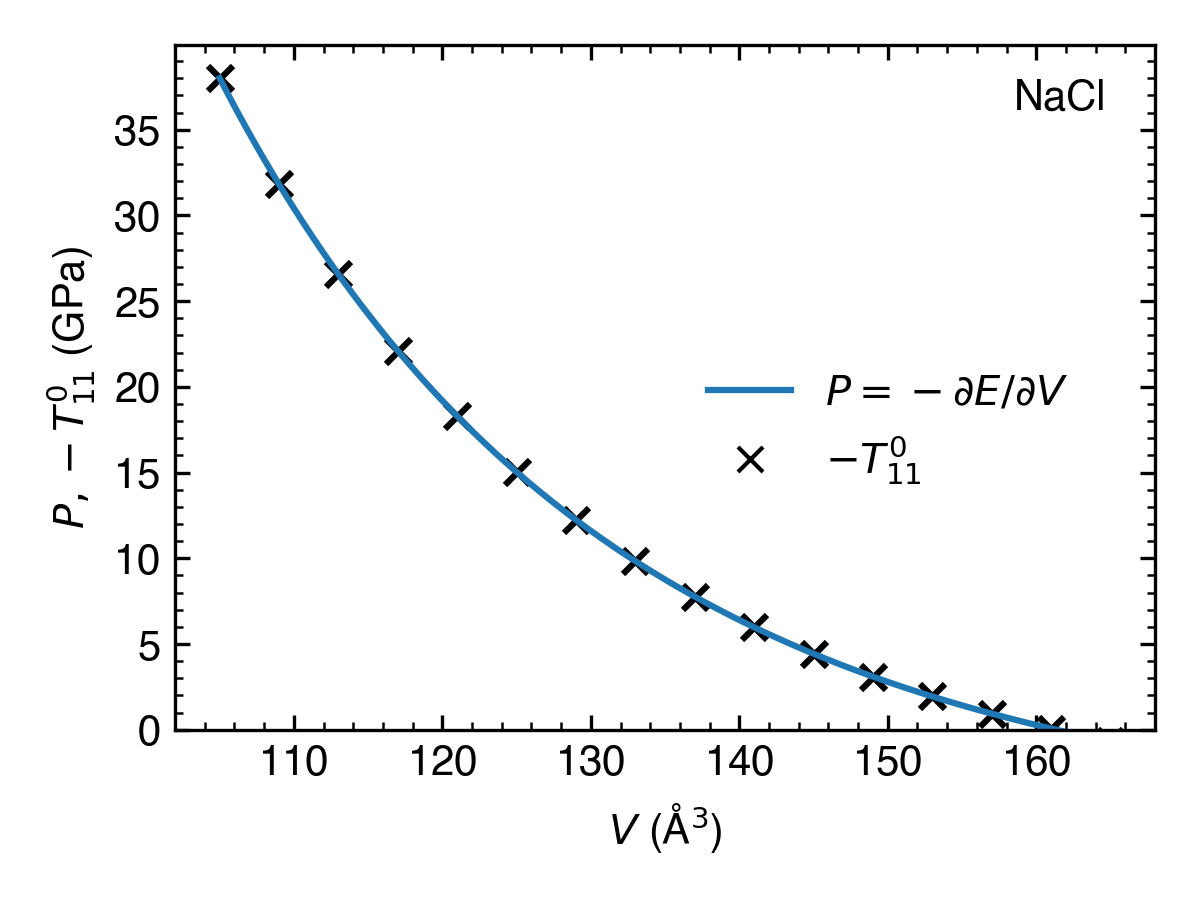}
\includegraphics[width=.4\textwidth]{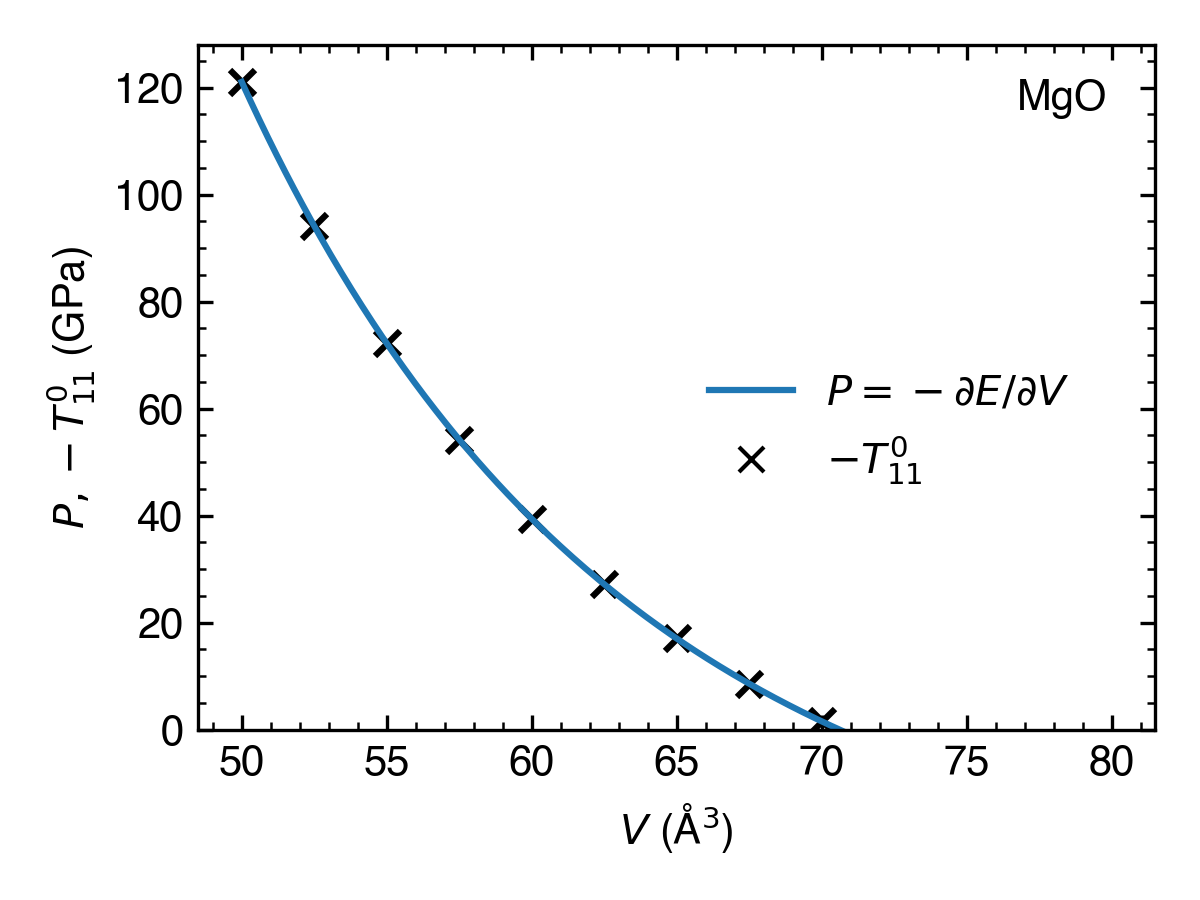}
\caption{Pressure vs.\ volume from third-order Birch–Murnaghan equation of state and $T^0_{11}$ from strain energy vs.\ strain expansion.}
\label{fig:s1}
\end{figure*}

\begin{figure*}[h]
\centering
\includegraphics[width=.4\textwidth]{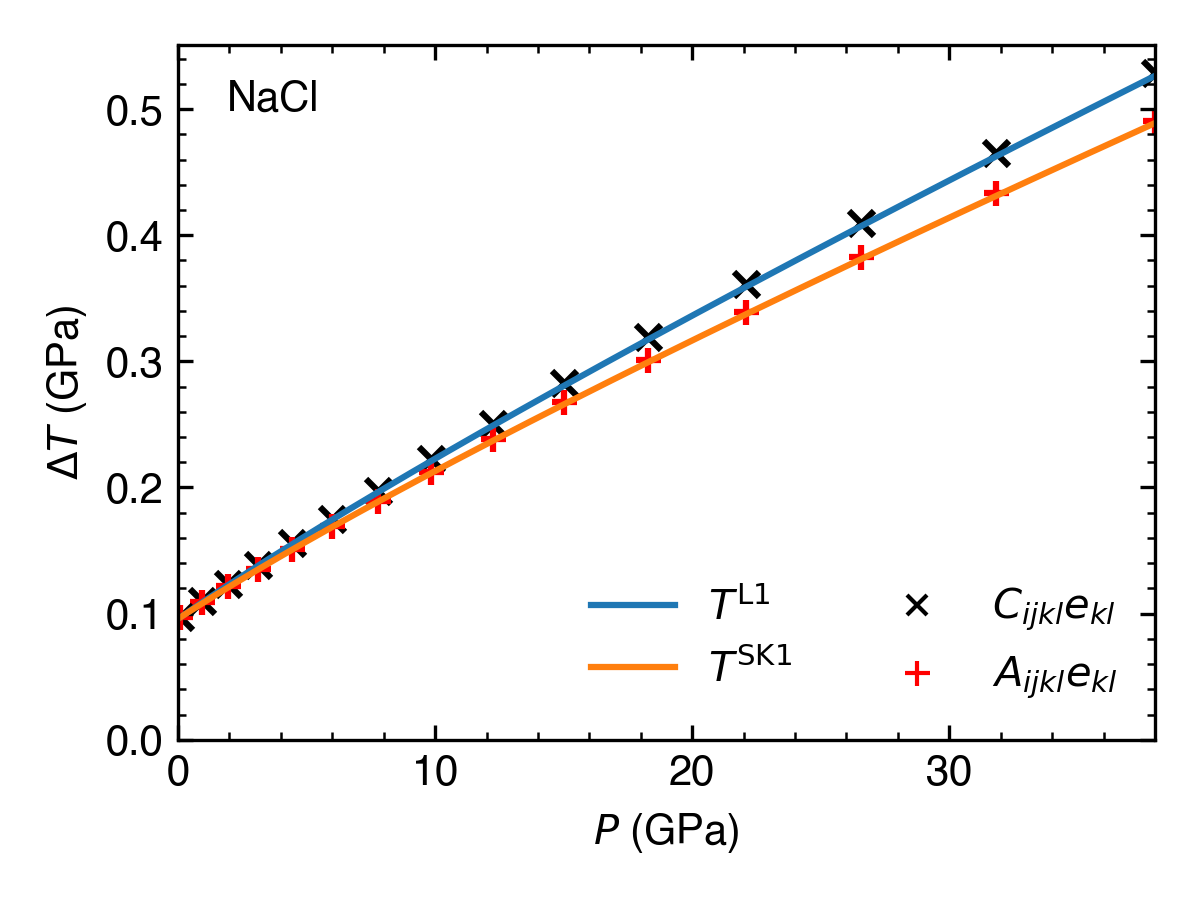}
\includegraphics[width=.4\textwidth]{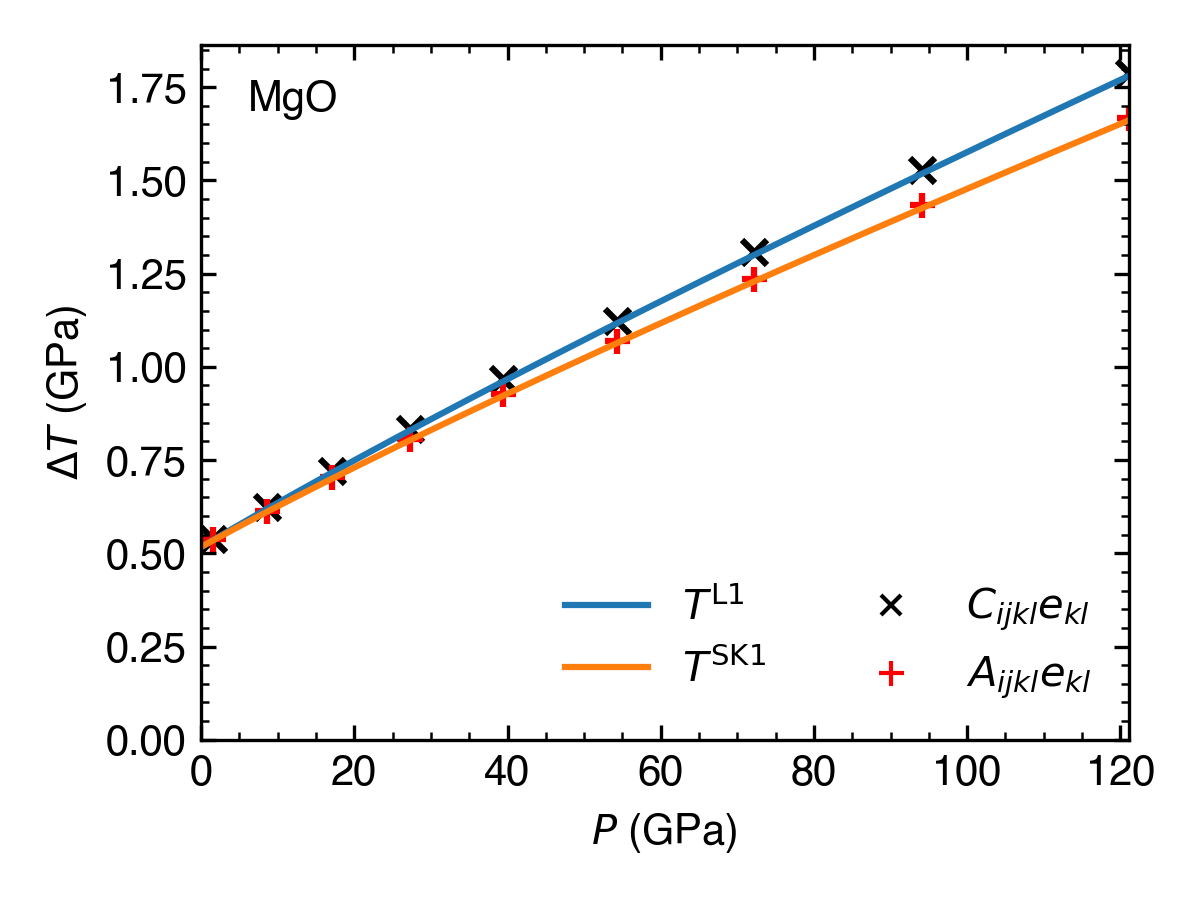}
\caption{Change in stress $T^\mathrm{L1}$ compared to $C_{ijkl}\, \eta_{kl}$ and $T^\mathrm{SK1}$ compared to $A_{ijkl}\, \eta_{kl}$ and $C_{ijkl} \,\eta_{kl}$ vs.\ $P$. A homogenous stretch $\eta_{11} = \eta_{22} = \eta_{33} = 0.001$ was applied.
The comparison outlines the difference between the incremental second Piola-Kirchoff stress $T^\mathrm{SK1}$ and the incremental Lagrangian Cauchy stress $T^\mathrm{L1}$. $T^\mathrm{SK1}$ is predicted by $A_{ijkl}$\,, with a hydrostatic initial stress, $T^\mathrm{L1}$ is predicted by $C_{ijkl}$\,. Here we calculate and compare these quantities in Fig.~S2 show these quantities w.r.t.\ a 0.1\% stretch.
}
\label{fig:s2}
\end{figure*}

\begin{figure*}[h]
\centering
\includegraphics[width=.4\textwidth]{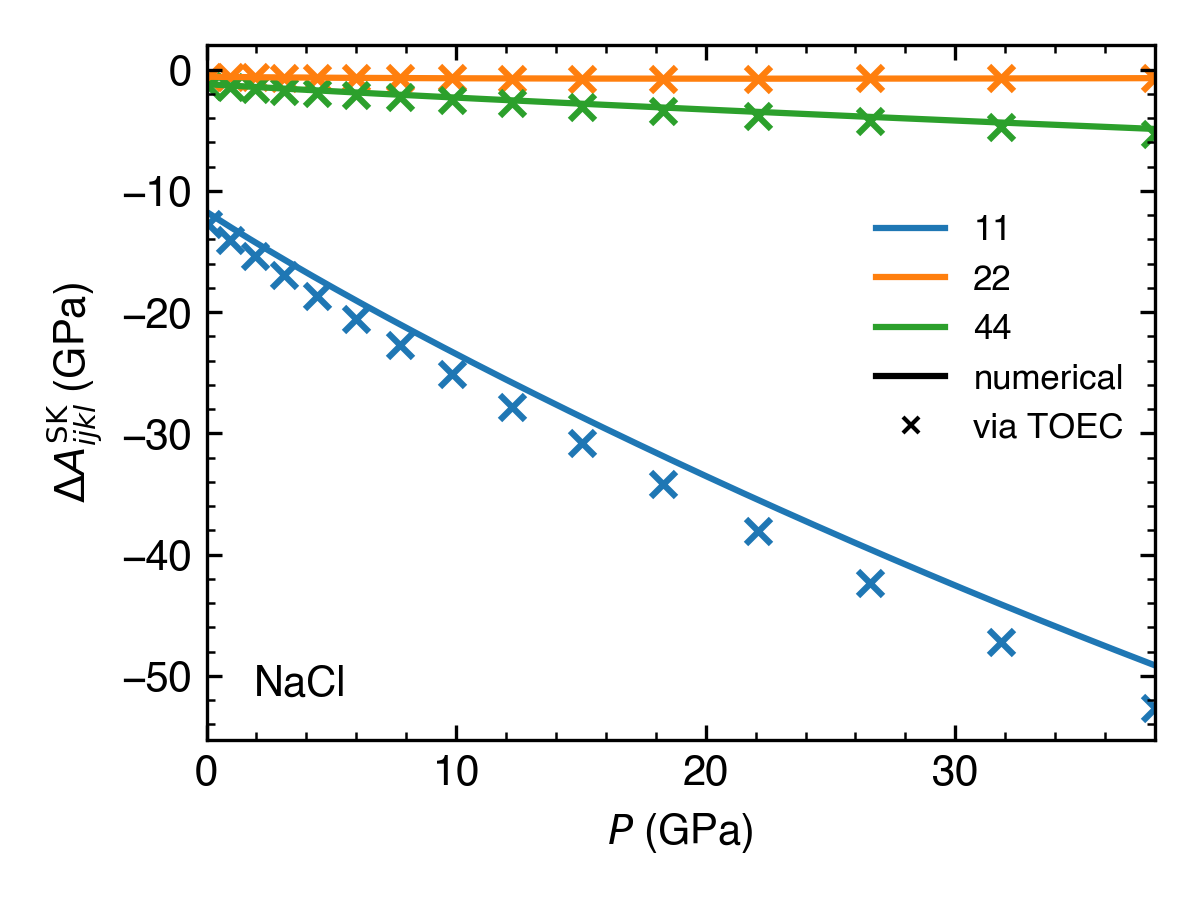}
\includegraphics[width=.4\textwidth]{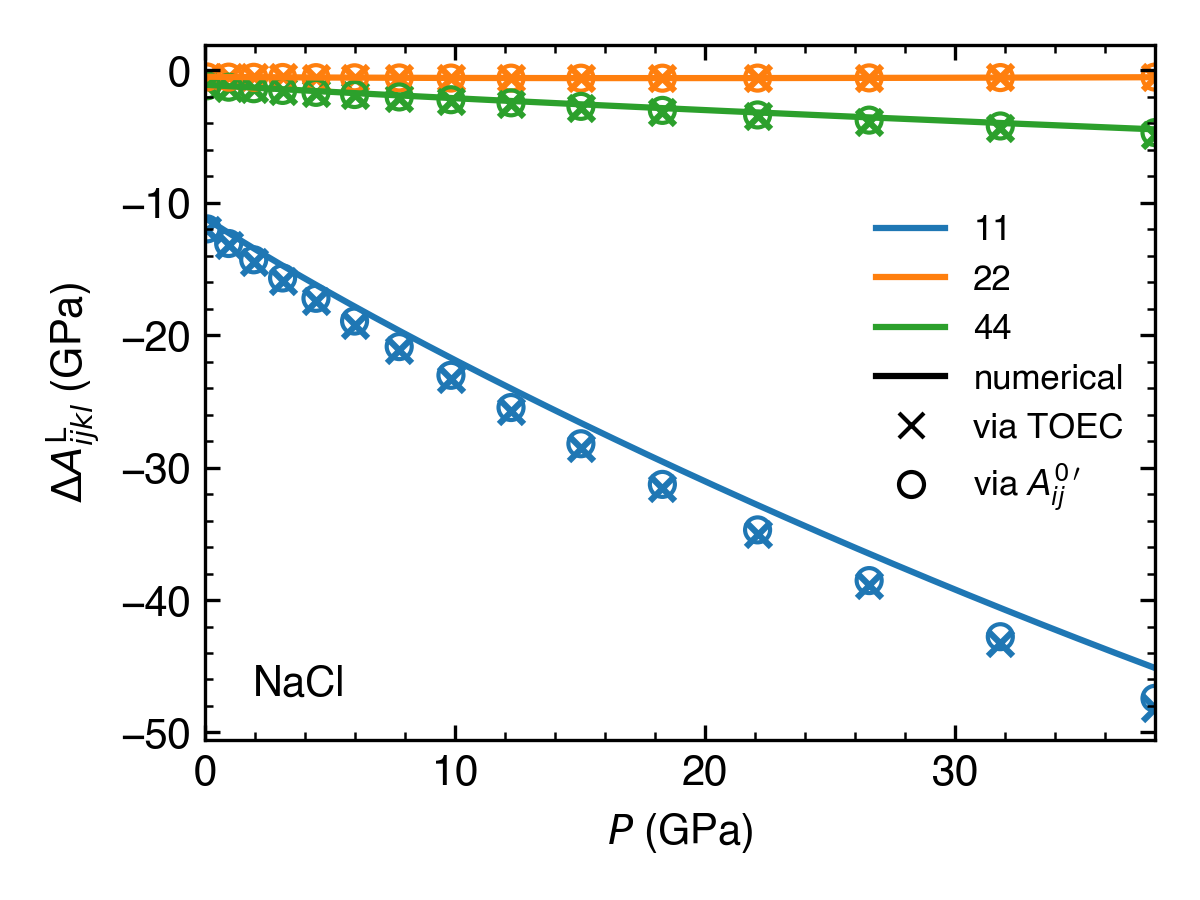}
\includegraphics[width=.4\textwidth]{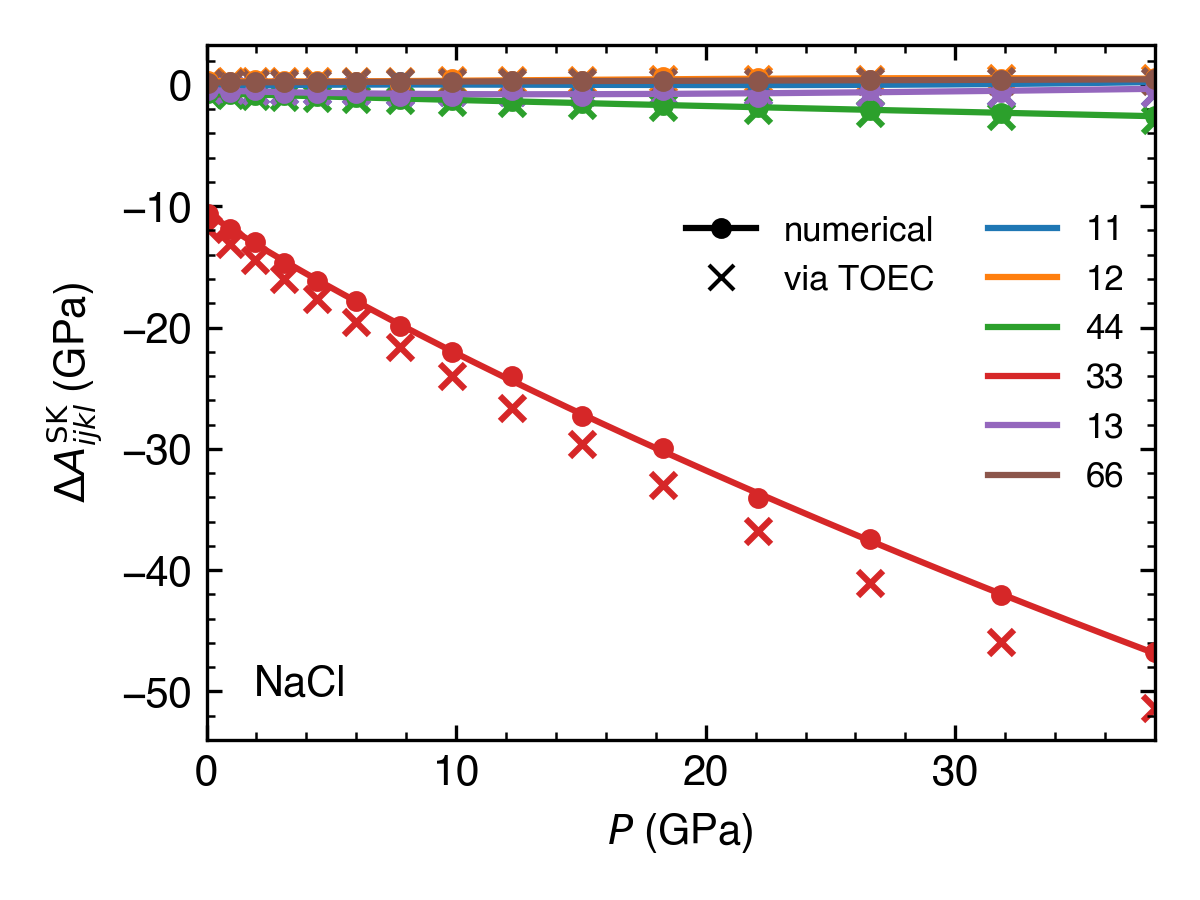}
\includegraphics[width=.4\textwidth]{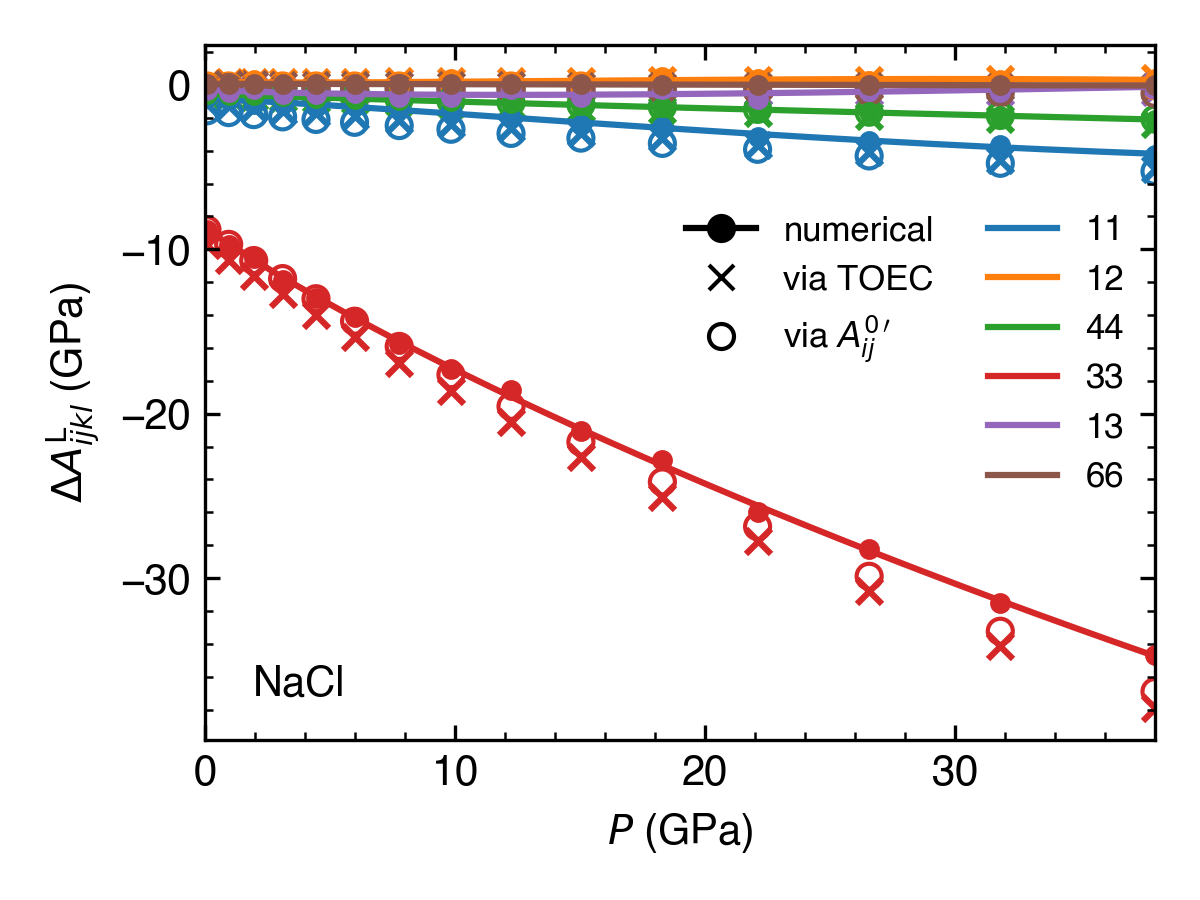}
\caption{Change in $A_{ijklmn}$ with a strain of 0.01 magnitude.}
\label{fig:s3}
\end{figure*}

